# Maximal Fisher Budget Forces Blind Directions in Bipartite Collective $SU(d)$ Metrology

Tariq Aziz[1], Naeem Akhtar[1], Dong Wang[1,*], Augusto Smerzi[2,†]

[1] *State Key Laboratory of Opto-Electronic Information Acquisition and Protection Technology, School of Physics, Anhui University, 230601, Hefei, China*

[2] *College of Engineering Physics, Shenzhen Technology University, Shenzhen 518118, China*

**Abstract**

For pure two-qudit probes used in a single fixed setting to estimate a common $SU(d)$ transformation, maximal total Fisher sensitivity and full local identifiability are mutually exclusive. We derive exact identities that determine the trace of the quantum Fisher information matrix from two measurable probe properties: exchange symmetry and collective polarization. Every maximizer is exchange symmetric and maximally entangled. Its Fisher matrix has equal sensitivity in every visible direction, while $d(d-1)/2$ generator directions are blind. More generally, weak compatibility, defined by vanishing mean generator commutators, forces at least $\lfloor d/2 \rfloor$ blind directions for every pure bipartite probe. A probe-defined antiunitary invariant classifies the complete Fisher spectrum at maximal entanglement, and approaches to the optimum that preserve exchange symmetry have a divergent Holevo cost. The bipartite obstruction is sharp in particle number: for $N \geq 3$, generalized GHZ probes can satisfy weak compatibility, attain the corresponding $N$-partite Fisher trace bound, and retain full local identifiability. Hence, Fisher sensitivity, weak compatibility, identifiability, and attainable precision are distinct resources in quantum metrology with noncommuting generators.

## I. Introduction

Complete multiparameter estimation of a unitary transformation generated by noncommuting observables requires more than maximizing total quantum Fisher information. The quantum Fisher information matrix (QFIM) quantifies local statistical distinguishability [1-14], and its trace in an orthonormal basis of $\mathfrak{su}(d)$ generators provides a basis-independent measure of total directional sensitivity [15, 16]. Estimating all parameters also requires local identifiability, equivalently a full-rank QFIM, while the attainable joint precision is governed by the Holevo bound rather than by the QFIM alone [5-14]. We ask whether a pure probe can maximize total Fisher sensitivity while remaining sensitive to every generator direction.

* dwang@ahu.edu.cn
† augustosmerzi@sztu.edu.cn

We answer this question for a pure probe of two $d$-level systems undergoing the same unknown collective transformation $U(\boldsymbol{\theta}) \otimes U(\boldsymbol{\theta})$, with $U(\boldsymbol{\theta}) \in SU(d)$. This is the smallest collective setting in which pure-state dimension counting does not preclude full local identifiability, since $\dim_{\mathbb{R}} \mathbb{CP}^{d^2-1} = 2d^2 - 2 > d^2 - 1 = \dim SU(d)$. Thus, any loss of rank is not forced by a shortage of state-space dimensions [17]. The collective encoding also differs fundamentally from ancilla-assisted estimation, where the unknown unitary acts on only one subsystem and maximally entangled probes can be optimal [18-23]. Previous work has established intrinsic Fisher-information sums [15,16], entanglement- and resource-dependent sensitivity bounds [24,25], and symmetry-induced blind directions [26]. Here we determine the exact obstruction for a pure bipartite probe used in one fixed collective setting.

Our first result is an exact no-go theorem that determines $\operatorname{tr} F$ from two measurable probe properties: exchange symmetry and collective polarization. It also fixes the complete maximizing set. Every maximizer is exchange symmetric and maximally entangled, but its QFIM is not full rank. Instead, $F_{\max} = \frac{16}{d} \Pi_{\mathfrak{p}}$, where $\Pi_{\mathfrak{p}}$ projects onto the $(d-1)(d+2)/2$-dimensional identifiable subspace. Every direction in this subspace carries the same Fisher information $16/d$, while the complementary $d(d-1)/2$ collective-generator directions are exactly blind. Hence, in a single fixed bipartite setting, maximal total Fisher sensitivity does not imply complete estimability: the optimum concentrates information on a proper subspace and removes access to an entire sector of collective parameters.

Our second result is an exclusion theorem for weak compatibility and local identifiability. For every pure bipartite probe in one fixed collective setting, $\Omega = 0 \Longrightarrow \operatorname{corank} F \geq \lfloor d/2 \rfloor$, where $\Omega$ is the matrix of mean generator commutators. Consequently, eliminating the commutator obstruction to simultaneous estimation necessarily creates blind directions. Away from exact compatibility, we bound the smallest Fisher eigenvalue by the incompatibility norm and the deviation of a reduced state from maximal mixing, showing that nonzero incompatibility can be structurally necessary to retain sensitivity to every parameter. The consequence is operational: along any fully identifiable, exchange-symmetric approach to the Fisher optimum, both the SLD benchmark and the attainable Holevo cost diverge. Compatibility, identifiability, and attainable precision are therefore distinct metrological resources.

We further solve the maximally entangled boundary by showing that an antiunitary invariant defined by the probe determines the complete Fisher spectrum. This classification also reveals that the Schmidt spectrum alone does not determine metrological performance: the relative Schmidt frame controls which generator directions become soft. Finally, the bipartite obstruction is sharp in particle number. For $N \geq 3$, generalized GHZ probes are weakly compatible, saturate the corresponding $N$-partite Fisher-trace bound, and possess a full-rank QFIM. This sharp change between two and three collectively driven subsystems shows that total Fisher sensitivity, weak compatibility, local identifiability, and attainable precision are distinct resources in multiparameter quantum metrology with noncommuting generators.

**II. Model and Fisher-budget identities**

We identify $\mathfrak{su}(d)$ with the real vector space of traceless Hermitian generators, and choose an orthonormal basis $\{g_\mu\}_{\mu=1}^{d^2-1}$, $\mathrm{Tr}(g_\mu g_\nu) = \delta_{\mu\nu}$. Throughout, Tr denotes a Hilbert-space trace, while tr denotes a trace over parameter-space matrices such as the QFIM. A pure bipartite probe is written as $|\psi_A\rangle = \sum_{i,j=1}^{d} A_{ij}\,|ij\rangle$, with $\mathrm{Tr}(AA^\dagger) = 1$. Define $U(\boldsymbol{\theta}) = \exp\big(-i\sum_\mu \theta_\mu g_\mu\big)$. The unknown collective transformation acts as $|\psi_A(\boldsymbol{\theta})\rangle = \big(U(\boldsymbol{\theta}) \otimes U(\boldsymbol{\theta})\big)|\psi_A\rangle$, with summation over repeated $\mu$. Equivalently, a generator $h \in \mathfrak{su}(d)$ produces the tangent vector $G_h|\psi_A\rangle = |\psi_{hA+Ah^\top}\rangle$, $G_h = h \otimes I_d + I_d \otimes h$, where $I_d$ is the identity on one $d$-dimensional subsystem. Therefore, a direction is invisible exactly when this tangent is vertical in projective Hilbert space

$$h \in \ker F \iff hA + Ah^\top = \mathrm{Tr}[(\rho_a + \rho_b)h]A, \tag{1}$$

where $\rho_a = AA^\dagger$, $\rho_b = (A^\dagger A)^\top$. Eq. (1) identifies the QFIM kernel with the stabilizer Lie algebra of the probe ray under collective $SU(d)$ action (Supplemental Material (SM) Sec. S3). The total Fisher budget is the trace of the QFIM over the orthonormal Lie-algebra basis. Let $\langle \mathcal{S} \rangle = \langle \psi_A|\mathcal{S}|\psi_A\rangle = \mathrm{Tr}(A^\dagger A^\top)$ be the exchange expectation, and define the collective polarization vector $b_\mu = \mathrm{Tr}\big[(\rho_a + \rho_b)g_\mu\big]$. A direct contraction of the pure-state QFIM gives (SM Sec. S2)

$$\mathrm{tr}\,F = 8\left(\frac{d^2-2}{d} + \langle \mathcal{S} \rangle\right) - 4\|b\|^2. \tag{2}$$

The trace has a direct isotropic interpretation. Let $m = d^2 - 1$, let $h(\mathbf{n}) = \sum_{\mu=1}^{m} n_\mu\, g_\mu$, with $\|\mathbf{n}\| = 1$, and let $d\nu(\mathbf{n})$ be normalized uniform measure on $S^{m-1}$. The one-direction QFI is

$$F_Q(\mathbf{n}) = \mathbf{n}^\top F \mathbf{n} \text{ and } \int_{S^{m-1}} \mathrm{F}_Q(\mathbf{n})\, d\nu(\mathbf{n}) = \frac{\mathrm{tr}\,F}{d^2-1}.$$

Thus, $\mathrm{tr}\, F$ is $d^2-1$ times the isotropic average one-direction sensitivity; it is not itself the attainable multiparameter cost. Hence, the scalar sensitivity budget is fixed by only two probe properties: exchange symmetry and collective polarization. Specifically, $\mathrm{tr}\, F \le \frac{8(d-1)(d+2)}{d}$, with equality if and only if $\langle \mathcal{S}\rangle = 1$ and $b = 0$. After converting the generator normalization used by Wilson et al. [16] to our convention $\mathrm{Tr}(g_\mu g_\nu) = \delta_{\mu\nu}$, their collective $SU(d)$ trace bound yields the same cap. Du et al. [24] obtained entanglement-dependent bounds for closely related sums of collective $SU(d)$ quantum Fisher information. The distinct content of Eq. (2) in the present pure bipartite model is an exact resolution rather than an inequality: the complete Fisher budget is expressed in terms of the exchange expectation $\langle \mathcal{S}\rangle$ and collective polarization $b$. Eq. (4) and Sec. III subsequently determine the complete equality set and show that every maximizer possesses a stabilizer-protected blind sector. The same polarization controls weak incompatibility. Define $\Omega_{\mu\nu} = \mathbb{I}\mathrm{m}[\langle \Delta G_\mu \Delta G_\nu \rangle] = -\frac{i}{2}\langle \psi_A | [G_\mu, G_\nu] | \psi_A \rangle$, where $\Delta G_\mu = G_\mu - \langle G_\mu \rangle$. The $SU(d)$ structure constants give $\|\Omega\|_{\mathrm{HS}}^2 = \frac{d}{2}\|b\|^2$. Combining this with Eq. (2) yields the exact sum rule

$$\mathrm{tr}\, F + \frac{8}{d}\|\Omega\|_{\mathrm{HS}}^2 = 8\left(\frac{d^2-2}{d} + \langle \mathcal{S}\rangle\right). \tag{3}$$

Unlike inequality-based trade-offs between Fisher information and measurement incompatibility [17, 27-29], Eq. (3) is an exact identity of the bipartite collective model. At fixed exchange symmetry, increasing the total Fisher budget is exactly equivalent to reducing weak incompatibility. Fig. 1(a) visualizes this conservation law: the upper dashed line is the exchange-symmetric Fisher cap, and the compatible edge already contains singular probes. Fig. 1(b) shows the complementary obstruction on the maximally entangled boundary: blind sectors are stratified by the antiunitary invariant $V = J^2$.

Finally, the distance from the Fisher-budget cap has an exact geometric form. Define $\delta = \rho_a + \rho_b - \frac{2I_d}{d}$, $\Delta = \frac{8(d-1)(d+2)}{d} - \mathrm{tr}\, F$. Since $\|b\|^2 = \|\delta\|_{\mathrm{HS}}^2$ and $\|A - A^\top\|_{\mathrm{HS}}^2 = 2 - 2\langle \mathcal{S}\rangle$, Eq. (2) gives

$$\Delta = 4\|A - A^\top\|_{\mathrm{HS}}^2 + 4\left\|\rho_a + \rho_b - \frac{2I_d}{d}\right\|_{\mathrm{HS}}^2. \tag{4}$$

The Fisher-budget cap is therefore reached precisely by probes that are exchange symmetric and have zero collective polarization. These two constraints force the maximizers onto a singular, stabilizer-protected orbit.

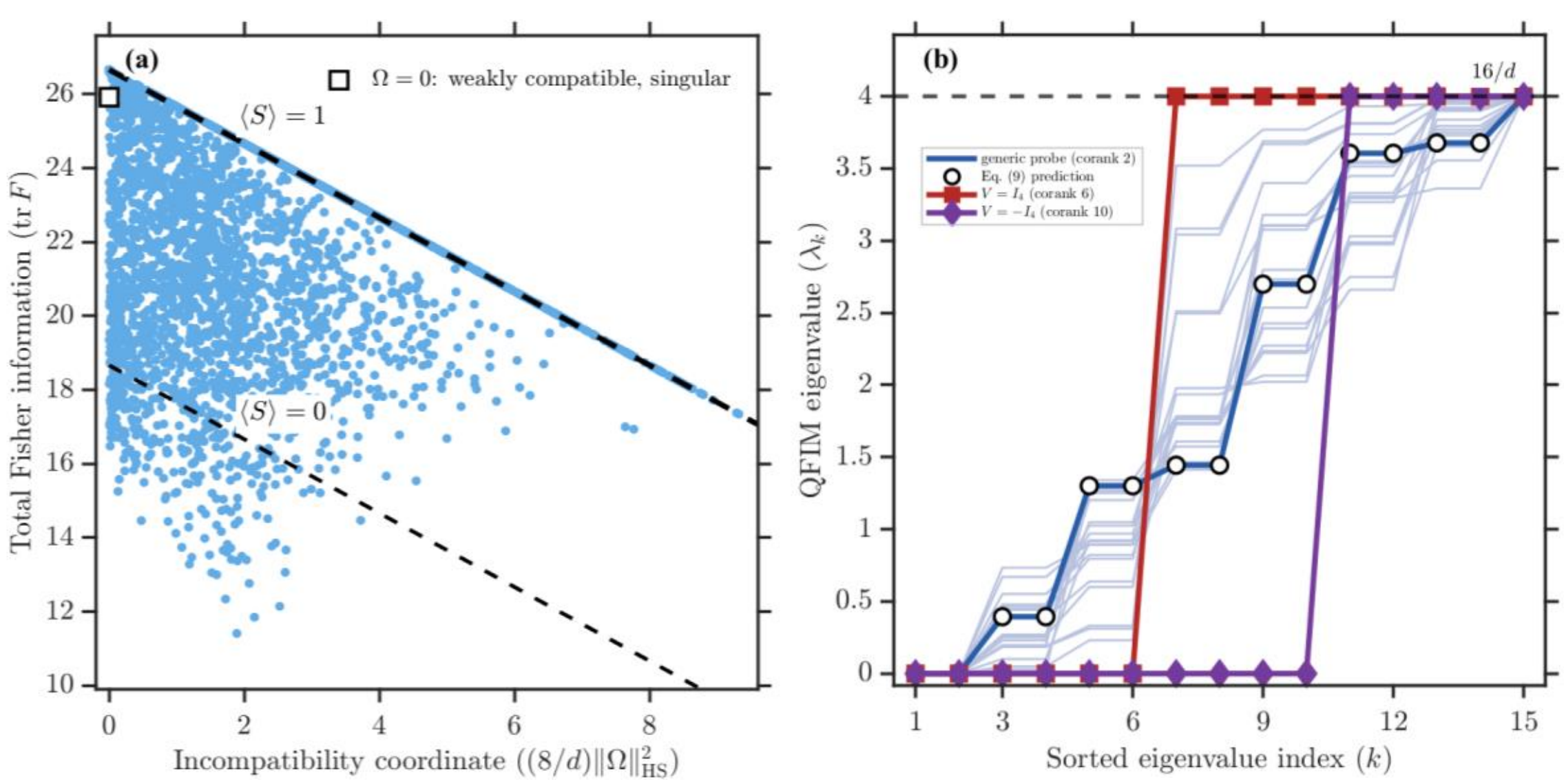

FIG. 1. (a) For $d = 3$, randomly generated pure probes obey the exact conservation law between $\mathrm{tr}\,F$ and $(8/d)\|\Omega\|_{\mathrm{HS}}^2$ [Eq. (3)]. The dashed lines mark the $\langle\mathcal{S}\rangle = 0$ sector and exchange-symmetric cap. The square marks a weakly compatible but singular nonmaximal probe. (b) Sorted QFIM spectra on the maximally entangled boundary for $d = 4$. Thin curves show generic boundary probes; the blue curve and open circles compare one representative spectrum with Eq. (9). Generic probes have corank two, whereas the $V = I_4$ maximizer and $V = -I_4$ stratum collapse to isotropic projector spectra with every visible eigenvalue equal to $16/d = 4$.

## III. Exact boundary spectrum and the projector optimum

A singular QFIM represents a nonregular statistical model with locally flat parameter directions [30-32]. In the present collective model, these flat directions are not accidental: Eq. (1) identifies them exactly with the infinitesimal stabilizer of the probe ray. The following results show how this stabilizer controls both the optimum and the full maximally entangled boundary.

Eq. (4) gives the rigidity of the Fisher-budget maximum. The cap is reached only if $A = A^{\top}$ and $\rho_a + \rho_b = \frac{2I_d}{d}$. For a complex symmetric $A$, the Autonne-Takagi factorization gives $A = uDu^{\top}$, with $u$ unitary and $D \geq 0$ diagonal [33]. Then $\rho_a = \rho_b = uD^2u^{\dagger}$, so $\rho_a + \rho_b = \frac{2I_d}{d}$ forces $D = I_d/\sqrt{d}$. Therefore, the Fisher-budget maximizers are exactly

$$A = \frac{uu^{\top}}{\sqrt{d}}, \quad u \in U(d), \tag{5}$$

namely the exchange-symmetric maximally entangled orbit. Hence the maximum of Eq. (2) is not attained on the full maximally entangled boundary, but only on a distinguished collective orbit. Since QFIM rank and spectrum are invariant under collective unitary conjugation, it is sufficient to evaluate the optimum at $A = I_d/\sqrt{d}$. For every traceless Hermitian generator $h$, $h^\top F h = \frac{4}{d}\|h + h^\top\|_{\mathrm{HS}}^2$. Under the Hilbert–Schmidt orthogonal decomposition $\mathfrak{su}(d) = \mathfrak{p} \oplus \mathfrak{so}(d)$, where $\mathfrak{p}$ denotes the real-symmetric traceless sector and $\mathfrak{so}(d)$ is represented by imaginary-antisymmetric Hermitian generators, this becomes

$$F = \frac{16}{d}\Pi_{\mathfrak{p}}. \tag{6}$$

Hence, the maximizer is isotropic on $\dim \mathfrak{p} = \frac{(d-1)(d+2)}{2}$ visible directions and blind on $\dim \mathfrak{so}(d) = \frac{d(d-1)}{2}$ stabilizer directions. The exact invariance $(O \otimes O)|\Phi\rangle = |\Phi\rangle$, $O \in O(d)$, identifies $SO(d)$ as the connected stabilizer of the maximizing probe: the optimum estimates the symmetric-space directions $SU(d)/SO(d)$ uniformly while deleting the isotropy directions exactly.

Antiunitary symmetries have previously been used to restore attainability of multiparameter bounds [34,35], while auxiliary entanglement can remove simultaneity constraints in $SU(2)$ estimation [36]. Here the antiunitary structure has a different role: it organizes the generator directions that are absent from the local statistical model.

The full maximally entangled boundary is organized by an antiunitary invariant. Write $|\psi_U\rangle = (I_d \otimes U)|\Phi\rangle$, $|\Phi\rangle = \frac{1}{\sqrt{d}}\sum_i |ii\rangle$, so that $A = \frac{U^\top}{\sqrt{d}}$. Then $b = 0$, and $\langle \mathcal{S}\rangle = \frac{\mathrm{Tr}(U\bar{U})}{d}$. Let $V = U\bar{U}$. The stabilizer condition Eq. (1) becomes

$$U\bar{h}U^\dagger = -h, \tag{7}$$

or equivalently, $\ker F$ is the $-1$ eigenspace of conjugation by the antiunitary $J = UC$, with $J^2 = V$, where $C$ is complex conjugation in the computational basis. Eq. (2) gives the boundary budget

$$\mathrm{tr}\, F_U = \frac{8}{d}(d^2 - 2 + \mathrm{Tr}\, V). \tag{8}$$

The boundary model is exactly solvable beyond its trace and kernel. Let $\mathcal{H}_d$ be the real Hilbert space of traceless Hermitian matrices with Hilbert-Schmidt inner product, and define $K_U(h) = U^\top \bar{h}\bar{U}$. The map $K_U\colon \mathcal{H}_d \to \mathcal{H}_d$ is real orthogonal, and $hA + Ah^\top = \frac{1}{\sqrt{d}}(h + K_U h)U^\top$. Because $b = 0$, the complete boundary QFIM is therefore

$$F_U = \frac{4}{d}(I + K_U)^*(I + K_U), \quad \operatorname{spec} F_U = \left\{\frac{8}{d}(1 + \cos\theta_k)\right\}_{k=1}^{d^2-1}, \tag{9}$$

where $*$ is the adjoint on $\mathcal{H}_d$, and $\theta_k$ are the canonical angles of the real-orthogonal map $K_U$, including 0 and $\pi$ for its $\pm$ sectors. Moreover, $K_U^2(h) = V^\dagger h V$, with $\operatorname{Tr}_{\mathbb{R}} K_U = \operatorname{Tr} V - 1$. Taking the trace of Eq. (9) therefore reproduces Eq. (8). Unitary-congruence classes of $U$ are fixed by the spectral class of $V$ [33,37]; consequently, the entire boundary QFIM spectrum, not only its corank, is a class function of $V$. The antiunitary frame fixes the orientation of the Fisher eigenspaces, whereas $V$ fixes their eigenvalues and multiplicities (SM Sec. S4). Eq. (9) gives the uniform boundary bound $0 \preccurlyeq F_U \preccurlyeq \frac{16}{d} I_{\mathcal{H}_d}$. At the extreme strata $V = I_d$ and $V = -I_d$, with the latter existing only for even $d$, one has $K_U^2 = I$. Therefore, every canonical angle is either 0 or $\pi$ and $F_U = \frac{16}{d}\Pi_U$, where $\Pi_U$ is the Hilbert-Schmidt orthogonal projector onto the identifiable sector. At the Fisher-budget optimum $V = I_d$,

$$\operatorname{spec} F = \left\{\left(\frac{16}{d}\right)^{\times(d-1)(d+2)/2}, \ 0^{\times d(d-1)/2}\right\}. \tag{10}$$

Hence, the optimum is not an anisotropic high-sensitivity metric. It is an isotropic projector: every visible collective direction carries exactly the same Fisher information $16/d$, while the stabilizer sector carries exactly none.

The blind directions also retain their Lie-algebraic structure. If $h, h' \in \ker F$, the corresponding collective generators stabilize the probe ray, and therefore their commutator does so as well; hence $[h, h']_{\mathfrak{g}} \equiv -i[h, h'] \in \ker F$. If $V$ has conjugate eigenvalue pairs $e^{\pm i\phi_j}$ of multiplicity $m_j$, together with $V = \pm 1$ eigenspaces of dimensions $n_\pm$, where $n_-$ is even, the threefold-way classification [38] gives

$$\operatorname{Ker} F \simeq \mathfrak{so}(n_+) \oplus \bigoplus_j \mathfrak{u}(m_j) \oplus \mathfrak{sp}\left(\frac{n_-}{2}\right)$$

and $\operatorname{corank} F = \frac{n_+(n_+-1)}{2} + \sum_j m_j^2 + \frac{n_-(n_-+1)}{2}$. Generic boundary probes therefore have the Abelian blind algebra $\mathfrak{u}(1)^{\oplus\lfloor d/2\rfloor}$; the optimum $V = I_d$ carries the algebra $\mathfrak{so}(d)$, which is non-Abelian for $d \geq 3$; and the $V = -I_d$ stratum for even $d$ carries $\mathfrak{sp}\left(\frac{d}{2}\right)$. Consequently, $\operatorname{corank} F \geq \lfloor d/2 \rfloor$ for every maximally entangled probe, with equality attained on the generic boundary stratum (SM Sec. S4). Budget and blindness are therefore nonmonotone: the Fisher-optimal probe is substantially more blind than a generic boundary probe, while for even $d$ the $V = -I_d$ stratum has an even larger corank $d(d+1)/2$.

**IV. Quantitative incompatibility-identifiability exclusion**

For a pure bipartite probe in one fixed collective setting, weak compatibility is the vanishing of the mean commutator matrix $\Omega_{\mu\nu}$. In the present model, the identity $\|\Omega\|_{\mathrm{HS}}^2 = (d/2)\|b\|^2$ gives, $\Omega = 0 \Leftrightarrow b = 0 \Leftrightarrow \rho_a + \rho_b = \frac{2I_d}{d}$. Therefore, weakly compatible bipartite probes are exactly the zero-collective-polarization probes. This condition already forces a kernel. Let $h_0 = \rho_a - \frac{I_d}{d}$, $\delta = \rho_a + \rho_b - \frac{2I_d}{d}$. An elementary coefficient-space identity gives $G_{h_0}|\psi_A\rangle = |\psi_{A\bar{\delta}}\rangle$ (SM Sec. S5). Hence, if $\Omega = 0$, then $\delta = 0$ and $G_{h_0}|\psi_A\rangle = 0$.

If $h_0 \neq 0$, this gives a nonzero element of $\ker F$. If $h_0 = 0$, then $\rho_a = I_d/d$, and the zero-polarization condition gives $\rho_b = I_d/d$; the probe is maximally entangled, so it is non-identifiable by the boundary theorem of Sec. III.

The complete compatible locus obeys a stronger universal bound. Diagonalizing $\rho_a$, compatibility implies that $\rho_b = \frac{2\,I_d}{d} - \rho_a$ is diagonal in the same basis, so $A$ maps each eigenspace $E_{2/d-\lambda}$ onto $E_\lambda$. The stabilizer equation therefore decomposes into complementary spectral pairs.

Each non-self-paired interior pair $0 < \lambda < 2/d$, $\lambda \neq 1/d$, of multiplicity $m$ contributes at least $m$ null generators through the Hermitian commutant of an $m \times m$ unitary block. The endpoint pair $(0, 2/d)$ contributes $m^2 \geq m$ null generators, while a self-paired block $\lambda = 1/d$ of dimension $r$ contributes at least $\lfloor r/2 \rfloor$ null directions by the boundary classification of Sec. III. The traces cancel within each complementary paired sector. Since $d = 2\sum_{\lambda<1/d} m_\lambda + r,$ one obtains (SM Sec. S5)

$$\Omega = 0 \implies \operatorname{corank} F \geq \left\lfloor \frac{d}{2} \right\rfloor. \tag{11}$$

The bound is sharp. For even $d$, choose $d/2$ one-dimensional complementary spectral pairs with generic paired-block phases; for odd $d$, use $(d-1)/2$ such pairs together with a single one-dimensional self-paired $1/d$ block. These constructions have exactly $\lfloor d/2 \rfloor$ null directions. Weak compatibility therefore defines a singular locus whose minimum possible corank is $\lfloor d/2 \rfloor$, whereas the Fisher-budget maximizer lies deeper, with $\operatorname{corank} F = d(d-1)/2$. Specifically, no pure bipartite probe under full collective $SU(d)$ encoding can simultaneously satisfy $\Omega = 0$ and possess a full-rank QFIM.

The exclusion has a quantitative extension away from exact compatibility. The same coefficient identity gives $h_0^\top F h_0 = 4\left[\mathrm{Tr}(\delta\rho_b\delta) - \left(\mathrm{Tr}(\delta h_0)\right)^2\right]$. Since $\mathrm{Tr}(\delta\rho_b\delta) \le \|\delta\|_{\mathrm{HS}}^2 = \frac{2}{d}\|\Omega\|_{\mathrm{HS}}^2$, the variational characterization of the smallest Fisher eigenvalue yields, whenever $\rho_a \neq I_d/d$,

$$\lambda_{\min}(F)\left\|\rho_a - \frac{I_d}{d}\right\|_{\mathrm{HS}}^2 \le \frac{8}{d}\|\Omega\|_{\mathrm{HS}}^2 \tag{12}$$

with the analogous bound for $\rho_b$ by exchanging the two subsystems (SM Sec. S5).

For every fully identifiable probe, Eq. (12) immediately gives the operational divergence certificate

$$C^{\mathrm{H}}(I) \ge C^{\mathrm{SLD}} = \mathrm{tr}\, F^{-1} \ge \frac{d\left\|\rho_a - \frac{I_d}{d}\right\|_{\mathrm{HS}}^2}{8\|\Omega\|_{\mathrm{HS}}^2}, \tag{12a}$$

see SM Eq. (S47). The bound need not determine the sharp divergence amplitude. However, it proves that at fixed nonzero local deviation of $\rho_a$ from maximal mixing, $\lambda_{\min}(F)$ is upper-bounded by a constant times $\|\Omega\|_{\mathrm{HS}}^2$, while the SLD cost diverges at least as $\|\Omega\|_{\mathrm{HS}}^{-2}$. Incompatibility is therefore not merely a measurement penalty in this model: a nonzero amount is required to maintain full local identifiability. Eq. (11) is the exactly compatible endpoint of this trade-off.

The quantities entering Eqs. (2)-(4), (12), and (12a) are probe-state observables: $b$ and the reduced states follow from one-body tomography, while $\langle\mathcal{S}\rangle$ is a swap expectation. The Fisher budget, its deficit, and the upper bound on the identifiability margin can therefore be determined without reconstructing the encoded process. The proof of the universal corank bound and constructions attaining it are given in SM Sec. S5.

**V. Attainable precision and multipartite sharpness**

The blind directions at the Fisher optimum have a direct precision cost. Throughout this section, $C^{\mathrm{H}}(I)$ denotes the identity-weight Holevo bound for local estimation at $\boldsymbol{\theta} = 0$ in the standard many-copy local-asymptotic setting [5,7,12]. On the exchange-symmetric manifold, the Takagi form $A = u\big(\mathrm{diag}(s_1, \dots, s_d)\big)u^\top$, $s_i = \sqrt{\lambda_i}$, $\lambda_i > 0$, $\sum_i \lambda_i = 1$, separates the collective generators into Cartan modes and symmetric and antisymmetric root modes (SM Sec. S6). For each pair $i < j$, the corresponding Fisher eigenvalues are $4\left(s_i + s_j\right)^2$ and $4\left(s_i - s_j\right)^2$. Hence, merging two Schmidt amplitudes makes an antisymmetric collective rotation unidentifiable. The Cartan sector remains regular as long as all $\lambda_i$ remain nonzero, while the exact exchange-symmetric Holevo bound is given in SM Sec. S6; for $d = 2$, it reduces, after normalization conversion, to the collective two-qubit result of Ref. [39].

The soft-mode structure yields a direct operational consequence. Define the Schmidt-overlap deficit $\Delta_* = d - \left(\sum_i \sqrt{\lambda_i}\right)^2 = \sum_{i<j}\left(s_i - s_j\right)^2$. Applying of the Cauchy-Schwarz inequality to the $d(d-1)/2$ antisymmetric root modes give (SM Sec. S7, Eq. (S63) and Eq. (S69)),

$$C^{\mathrm{H}}(I) \geq C^{\mathrm{SLD}} = \mathrm{tr}\, F^{-1} \geq \frac{[d(d-1)/2]^2}{4\Delta_*} \underset{\Delta\to 0}{\geq} \frac{4(d-1)^2}{\Delta}[1 + o(1)]. \tag{13}$$

where $\Delta$ is the Fisher-budget deficit defined in Eq. (4) (SM Sec. S7). Hence, every fully identifiable exchange-symmetric approach to the optimum has a Holevo cost that diverges at least inversely with the lost Fisher budget. The approach increases sensitivity in the visible sector while driving the emerging stabilizer directions to infinite estimation cost.

The antiunitary boundary class determines how many such directions become soft. The two qutrit paths in SM Sec. S11 have the same Schmidt spectrum but approach different $V$ classes. Along the exchange-symmetric path, three Fisher eigenvalues vanish quadratically; along a rotated path, only one does. Their leading Holevo costs differ by a factor of nine. The relative Schmidt frame therefore controls attainable precision independently of the Schmidt values and hence of the amount of entanglement.

The obstruction is sharp in particle number. For $N$ collectively driven qudits, the Fisher-budget cap is $4N(d-1)(d+N)/d$. For $N \geq 3$, the generalized GHZ probe satisfies

$$\left|\mathrm{GHZ}_{d,N}\right\rangle = \frac{1}{\sqrt{d}}\sum_{a=1}^{d}|\mathrm{a}\rangle^{\otimes N}, \quad F_{\mathrm{GHZ}_{d,N}} = \frac{4N}{d} I_{\mathrm{off}} \oplus \frac{4N^2}{d} I_C, \tag{14}$$

where $I_{\mathrm{off}}$ and $I_C$ denote the identity operators on the off-diagonal and Cartan (diagonal traceless) generator subspaces, acting on the $d(d-1)$ off-diagonal generators and the $d-1$ Cartan generators, respectively (SM Sec. S12). The state definition and the exact QFIM decomposition are derived in SM Sec. S12, Eqs. (S101) and (S104), respectively. This probe is permutation symmetric, has zero collective polarization, and saturates the $N$-partite Fisher cap. Moreover, $\lambda_{\min}(F) = 4N/d > 0$. Therefore, it remains fully identifiable while being weakly compatible and Fisher-budget maximizing.

Fig. 2(a) and 2(b) illustrates both consequences: the inverse divergence of the estimation cost near the Fisher optimum and the ninefold difference between equal-Schmidt-spectrum paths approaching distinct antiunitary boundary classes.

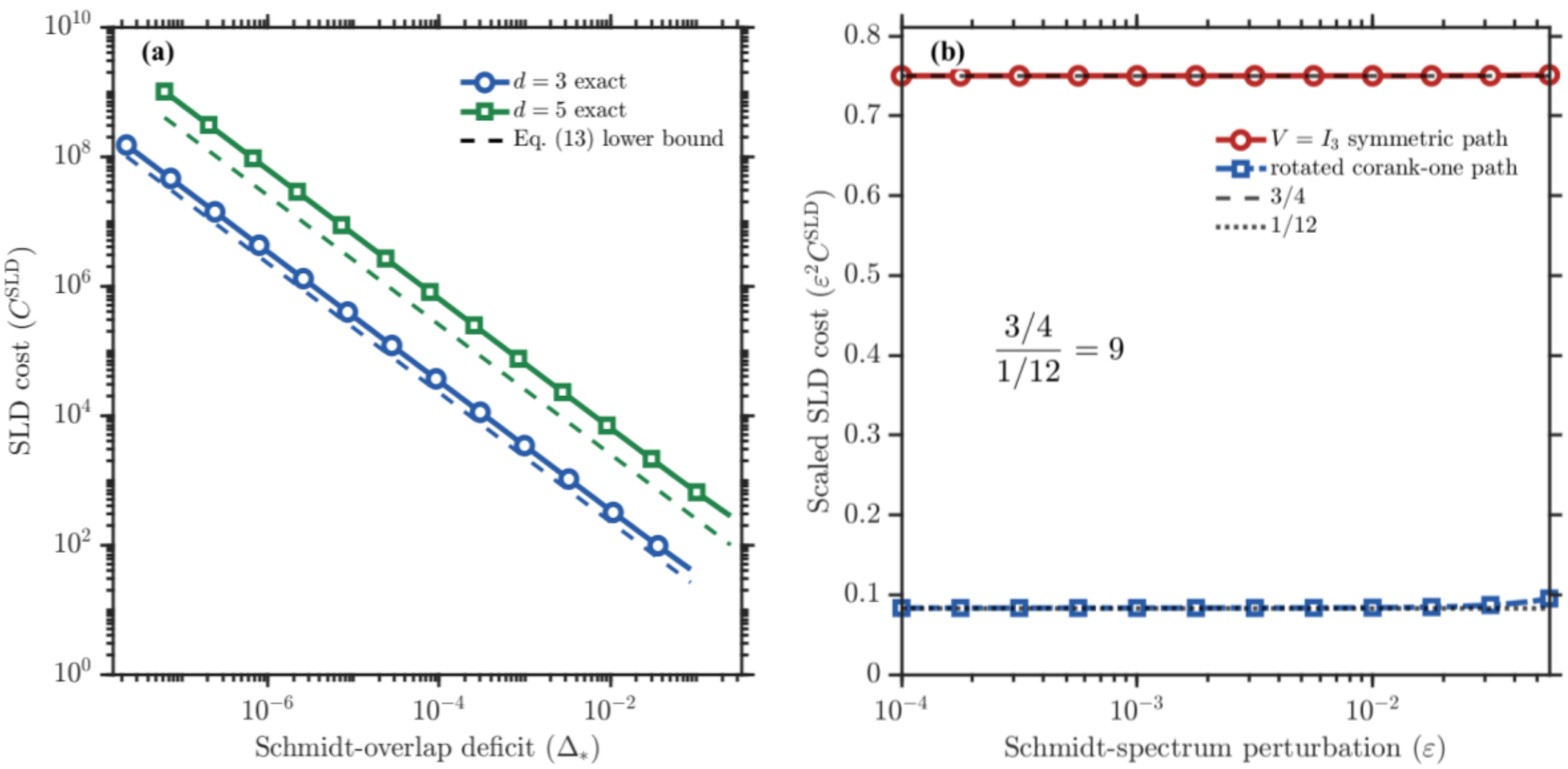

FIG. 2. (a) For exchange-symmetric probes, the SLD cost diverges inversely in the Schmidt-overlap deficit $\Delta_*$. Solid curves are exact values; dashed curves show the lower bound in Eq. (13). (b) For qutrit probes with the same Schmidt spectrum, the scaled costs $\varepsilon^2 C^{\mathrm{SLD}}$ approach $3/4$ along the exchange-symmetric path and $1/12$ along a rotated corank-one path, giving a ninefold separation. The corresponding Holevo costs have the same leading coefficients because the Holevo correction is subleading along these paths.

The particle-number threshold is exact. For $N = 1$, the pure-state orbit has real dimension at most $2d - 2 < d^2 - 1$, excluding complete $SU(d)$ identification. For $N = 2$, Eq. (11) proves that no weakly compatible pure probe is fully identifiable, and every Fisher-budget maximizer has $d(d-1)/2$ blind directions. Eq. (14) establishes the converse for every $N \geq 3$. Therefore, a weakly compatible, Fisher-budget-maximizing, fully identifiable pure collective probe exists if and only if $N \geq 3$.

As a supplementary asymptotic result, the optimized identity-weight SLD benchmark over fully identifiable pure bipartite probes satisfies $C_*^{\mathrm{SLD}}(d) = \Omega(d^3)$ and $C_*^{\mathrm{SLD}}(d) = O(d^3 \log d)$, whereas fully identifiable exchange-symmetric probes obey an $\Omega(d^4)$ lower bound (SM Secs. S13-S15). These asymptotic bounds do not enter the exact bipartite no-go theorem or the particle-number threshold.

The blind-sector theorem is thus a sharp consequence of collective action on exactly two subsystems, not of collective $SU(d)$ covariance itself. Eqs. (6), (9), (11), (12a), (13), and (14) give a single structural picture. At the bipartite optimum, symmetry concentrates the available Fisher budget isotropically on a proper subspace and deletes the complementary directions. Weak compatibility enforces a universal blindness floor throughout the bipartite model. Additional subsystems can restore the missing tangent directions and permit compatibility, maximality, and full identifiability to coexist.

## VI. Conclusion

Bipartite collective $SU(d)$ metrology separates scalar Fisher strength from estimability. Exact budget and deficit identities force every Fisher-budget maximizer onto the exchange-symmetric maximally entangled orbit. On the full maximally entangled boundary, the QFIM spectrum is a class function of the antiunitary invariant $V = U\bar{U}$. At the optimum it collapses to the isotropic projector $F = \frac{16}{d}\Pi_{\mathrm{p}}$. Therefore, every visible generator carries identical Fisher information and every stabilizer generator is exactly blind. Geometrically, this symmetry-controlled loss of rank provides a parameter-space analogue of codimension-controlled QFI singularities at topological band-touching defects [40].

The same obstruction persists beyond the optimum. Weak compatibility enforces at least $\lfloor d/2 \rfloor$ blind directions for every pure bipartite probe, while Eqs. (12) and (12a) quantitatively bound the identifiability margin and attainable cost by the incompatibility norm. The singularity is operational: exchange-symmetric approaches to the cap incur an inverse-deficit Holevo divergence, and the antiunitary class controls how many precision channels become soft. However, the obstruction ends sharply at two subsystems. For $N \geq 3$, generalized GHZ probes are weakly compatible, Fisher-budget maximizing, and fully identifiable. The no-go theorem concerns one fixed probe setting; it does not exclude experimental designs that combine several known probe orientations. Fisher strength, weak compatibility, attainable precision, and identifiability are therefore distinct resources; maximizing one can project away another.

## ACKNOWLEDGMENTS

This work was supported by the National Natural Science Foundation of China (Grants No. 12475009 and No. 12075001), Anhui Provincial Key Research and Development Plan (Grant No. 2022b13020004), Anhui Province Science and Technology Innovation Project (Grant No. 202423r06050004), the Anhui Provincial Natural Science Foundation (Grant No. 2508085ZD001), Anhui Provincial Department of Industry and Information Technology (Grant No. JB24044), and Anhui Provincial University Scientific Research Major Project (Grant No. 2024AH040008). T.A.

would like to acknowledge Mirpur University of Science and Technology (MUST), Mirpur, AJ&K, Pakistan, for granting ex-country leave. A.S. acknowledges support from the National Natural Science Foundation of China under Grant No. W2531008 and from the Peacock Plan.

**References**

[1] C. W. Helstrom, *Quantum Detection and Estimation Theory* (Academic Press, New York, 1976).

[2] S. L. Braunstein and C. M. Caves, Statistical distance and the geometry of quantum states, Phys. Rev. Lett. **72**, 3439 (1994).

[3] M. G. A. Paris, Quantum estimation for quantum technology, Int. J. Quantum Inf. **7**, 125 (2009).

[4] L. Pezzè and A. Smerzi, Entanglement, nonlinear dynamics, and the Heisenberg limit, Phys. Rev. Lett. **102**, 100401 (2009).

[5] A. S. Holevo, *Probabilistic and Statistical Aspects of Quantum Theory*, 2nd ed. (Edizioni della Normale, Pisa, 2011).

[6] A. Fujiwara and H. Nagaoka, Quantum Fisher metric and estimation for pure state models, Phys. Lett. A **201**, 119 (1995).

[7] K. Matsumoto, A new approach to the Cramér-Rao-type bound of the pure-state model, J. Phys. A **35**, 3111 (2002).

[8] F. Albarelli, J. F. Friel, and A. Datta, Evaluating the Holevo Cramér-Rao bound for multiparameter quantum metrology, Phys. Rev. Lett. **123**, 200503 (2019).

[9] M. Szczykulska, T. Baumgratz, and A. Datta, Multi-parameter quantum metrology, Adv. Phys. X **1**, 621 (2016).

[10] S. Ragy, M. Jarzyna, and R. Demkowicz-Dobrzański, Compatibility in multiparameter quantum metrology, Phys. Rev. A **94**, 052108 (2016).

[11] J. Liu, H. Yuan, X.-M. Lu, and X. Wang, Quantum Fisher information matrix and multiparameter estimation, J. Phys. A **53**, 023001 (2020).

[12] R. Demkowicz-Dobrzański, W. Górecki, and M. Guţă, Multi-parameter estimation beyond quantum Fisher information, J. Phys. A **53**, 363001 (2020).

[13] J. S. Sidhu, Y. Ouyang, E. T. Campbell, and P. Kok, Tight bounds on the simultaneous estimation of incompatible parameters, Phys. Rev. X **11**, 011028 (2021).

[14] L. Pezzè, M. A. Ciampini, N. Spagnolo, P. C. Humphreys, A. Datta, I. A. Walmsley, M. Barbieri, F. Sciarrino, and A. Smerzi, Optimal measurements for simultaneous quantum estimation of multiple phases, Phys. Rev. Lett. **119**, 130504 (2017).

[15] A. Z. Goldberg, L. L. Sánchez-Soto, and H. Ferretti, Intrinsic sensitivity limits for multiparameter quantum metrology, Phys. Rev. Lett. **127**, 110501 (2021).

[16] C. Wilson, J. D. Wilson, L. Coffman, S. S. Alam, and M. J. Holland, Geometric invariants of quantum metrology, Phys. Rev. A **113**, 063725 (2026).

[17] A. Candeloro, Z. Pazhotan, and M. G. A. Paris, Dimension matters: precision and incompatibility in multi-parameter quantum estimation models, Quantum Sci. Technol. **9**, 045045 (2024).

[18] A. Fujiwara, Quantum channel identification problem, Phys. Rev. A **63**, 042304 (2001).

[19] A. Acín, E. Jané, and G. Vidal, Optimal estimation of quantum dynamics, Phys. Rev. A **64**, 050302(R) (2001).

[20] M. A. Ballester, Estimation of unitary quantum operations, Phys. Rev. A **69**, 022303 (2004).

[21] G. Chiribella, G. M. D'Ariano, and M. F. Sacchi, Optimal estimation of group transformations using entanglement, Phys. Rev. A **72**, 042338 (2005).

[22] J. Kahn, Fast rate estimation of a unitary operation in $SU(d)$, Phys. Rev. A **75**, 022326 (2007).

[23] A. Fujiwara, Estimation of $SU(2)$ operation and dense coding: An information geometric approach, Phys. Rev. A **65**, 012316 (2001).

[24] S. Du, S. Liu, M. Fadel, G. Vitagliano, and Q. He, Uncertainty relations between quantum Fisher information and entanglement monotones, Phys. Rev. Lett. **136**, 110806 (2026).

[25] S. Du, S. Liu, F. E. S. Steinhoff, and G. Vitagliano, Characterizing resources for multiparameter estimation of $SU(2)$ and $SU(1,1)$ unitaries, Quantum **10**, 2130 (2026).

[26] A. Z. Goldberg, A. B. Klimov, G. Leuchs, and L. L. Sánchez-Soto, Rotation sensing at the ultimate limit, J. Phys. Photonics **3**, 022008 (2021).

[27] A. Carollo, B. Spagnolo, A. A. Dubkov, and D. Valenti, On quantumness in multi-parameter quantum estimation, J. Stat. Mech. 094010 (2019).

[28] X.-M. Lu and X. Wang, Incorporating Heisenberg's uncertainty principle into quantum multiparameter estimation, Phys. Rev. Lett. **126**, 120503 (2021).

[29] F. Belliardo and V. Giovannetti, Incompatibility in quantum parameter estimation, New J. Phys. **23**, 063055 (2021).

[30] D. Šafránek, Discontinuities of the quantum Fisher information and the Bures metric, Phys. Rev. A **95**, 052320 (2017).

[31] A. Z. Goldberg, J. L. Romero, Á. S. Sanz, and L. L. Sánchez-Soto, Taming singularities of the quantum Fisher information, Int. J. Quantum Inf. **19**, 2140004 (2021).

[32] G. Mihailescu, S. Sarkar, A. Bayat, S. Campbell, and A. K. Mitchell, Metrological symmetries in singular quantum multi-parameter estimation, Quantum Sci. Technol. **11**, 015006 (2026).

[33] R. A. Horn and C. R. Johnson, *Matrix Analysis*, 2nd ed. (Cambridge University Press, Cambridge, 2013).

[34] J. Miyazaki and K. Matsumoto, Imaginarity-free quantum multiparameter estimation, Quantum **6**, 665 (2022).

[35] B. Wang, K. Zheng, Q. Xie, A. Zhang, L. Xu, and L. Zhang, Achieving the multiparameter quantum Cramér-Rao bound with antiunitary symmetry, Phys. Rev. Lett. **133**, 210801 (2024).

[36] Y. Yang, S. Ru, M. An, Y. Wang, F. Wang, P. Zhang, and F. Li, Multiparameter simultaneous optimal estimation with an $SU(2)$ coding unitary evolution, Phys. Rev. A **105**, 022406 (2022).

[37] R. A. Horn and V. V. Sergeichuk, Canonical forms for unitary congruence and *congruence, Linear Multilinear Algebra **57**, 777 (2009).

[38] F. J. Dyson, The threefold way. Algebraic structure of symmetry groups and ensembles in quantum mechanics, J. Math. Phys. **3**, 1199 (1962).

[39] J. Friel, P. Palittapongarnpim, F. Albarelli, and A. Datta, Attainability of the Holevo-Cramér-Rao bound for two-qubit 3D magnetometry, arXiv:2008.01502 [quant-ph] (2020).

[40] C. A. S. Almeida, Codimension-controlled universality of quantum Fisher information singularities at topological band-touching defects, arXiv:2604.01515 [quant-ph] (2026).

# Supplemental Material

## S1. Conventions and completeness identities

The bipartite probe and normalization are

$$|\psi_A\rangle = \sum_{i,j=1}^{d} A_{ij}|ij\rangle, \quad \mathrm{Tr}(AA^\dagger) = 1. \tag{S1}$$

Here $A \in M_d(\mathbb{C})$ is the coefficient matrix of the bipartite pure state in the computational product basis. For single-system matrices $X, Y$,

$$(X \otimes Y)|\psi_A\rangle = |\psi_{XAY^\top}\rangle, \quad \langle\psi_A|X \otimes Y|\psi_A\rangle = \mathrm{Tr}(A^\dagger \mathrm{X} \mathrm{A} Y^\top). \tag{S2}$$

With $|\Phi\rangle = d^{-1/2} \sum_i |ii\rangle$, we use

$$(M \otimes I_d)|\Phi\rangle = (I_d \otimes M^\top)|\Phi\rangle. \tag{S3}$$

The generators $\{g_\mu\}_{\mu=1}^{d^2-1}$ are Hermitian, traceless, and orthonormal,

$$\mathrm{Tr}(g_\mu g_\nu) = \delta_{\mu\nu}, \tag{S4}$$

They obey the completeness identity

$$\sum_\mu (g_\mu)_{ab}(g_\mu)_{cd} = \delta_{ad}\delta_{bc} - \frac{1}{d}\delta_{ab}\delta_{cd}, \tag{S5}$$

consequently

$$\sum_\mu g_\mu^2 = \frac{d^2-1}{d} I_d, \quad \sum_\mu g_\mu \otimes g_\mu = \mathcal{S} - \frac{1}{d} I_d \otimes I_d, \tag{S6}$$

where $\mathcal{S}$ is the swap operator $\mathcal{S}|ij\rangle = |ji\rangle$. Writing $[g_\mu, g_\nu] = i \sum_\lambda f_{\mu\nu\lambda} g_\lambda$ with real structure constants $f_{\mu\nu\lambda}$, the adjoint Casimir in this normalization is

$$\sum_{\mu,\nu} f_{\mu\nu\lambda} f_{\mu\nu\kappa} = 2d\delta_{\lambda\kappa}. \tag{S7}$$

Throughout, Tr denotes Hilbert-space trace, while tr denotes trace over parameter-space matrices such as the QFIM.

## S2. Fisher-budget identity and curvature relation

Let $G_\mu = g_\mu \otimes I_d + I_d \otimes g_\mu$ be the collective bipartite generator and $b = (b_\mu)_{\mu=1}^{d^2-1}$ be its collective polarization vector, defined by $b_\mu = \langle G_\mu \rangle = \mathrm{Tr}[(\rho_a + \rho_b) g_\mu]$, $\rho_a = AA^\dagger$, $\rho_b = (A^\dagger A)^\top$. For a pure unitary model, $F_{\mu\nu} = 4\mathbb{R}\mathrm{e}(\langle \Delta G_\mu \Delta G_\nu \rangle)$, $\Delta G_\mu = G_\mu - \langle G_\mu \rangle$. Therefore,

$$\mathrm{tr}\, F = 4 \sum_\mu (\langle G_\mu^2 \rangle - b_\mu^2). \tag{S8}$$

Since

$$G_\mu^2 = g_\mu^2 \otimes I_d + I_d \otimes g_\mu^2 + 2 g_\mu \otimes g_\mu, \tag{S9}$$

Eq. (S6) gives

$$\sum_{\mu} \langle G_\mu^2 \rangle = 2\left(\frac{d^2-2}{d} + \langle \mathcal{S} \rangle\right), \tag{S10}$$

where $\langle \mathcal{S} \rangle = \langle \psi_A | \mathcal{S} | \psi_A \rangle = \mathrm{Tr}(A^\dagger A^\top)$. Substitution in Eq. (S8) yields

$$\mathrm{tr}\, F = 8\left(\frac{d^2-2}{d} + \langle \mathcal{S} \rangle\right) - 4\|b\|^2. \tag{S11}$$

This is Eq. (2) of the main text. Let $m = d^2 - 1$, let $h(\mathbf{n}) = \sum_{\mu=1}^{m} n_\mu\, g_\mu$, with $\|\mathbf{n}\| = 1$, and let $d\nu(\mathbf{n})$ be normalized uniform measure on $S^{m-1}$. The one-direction QFI is $F_Q(\mathbf{n}) = \mathbf{n}^\top F \mathbf{n}$ and $\int_{S^{m-1}} \mathrm{F}_Q(\mathbf{n})\, d\nu(\mathbf{n}) = \frac{\mathrm{tr}\, F}{d^2-1}$. Thus, $\mathrm{tr}\, F$ is $d^2 - 1$ times the isotropic average one-direction sensitivity; it is not itself the attainable multiparameter cost. The imaginary part of the pure-state quantum geometric tensor is

$$\Omega_{\mu\nu} \coloneqq \mathbb{I}m\, Q_{\mu\nu} = -\frac{i}{2}\langle [G_\mu, G_\nu] \rangle, \tag{S12}$$

where $Q_{\mu\nu} = \langle \Delta G_\mu \Delta G_\nu \rangle$. Because $[G_\mu, G_\nu] = i \sum_\lambda f_{\mu\nu\lambda} G_\lambda$, we have

$$\Omega_{\mu\nu} = \frac{1}{2}\sum_{\lambda} f_{\mu\nu\lambda} b_\lambda \tag{S13}$$

Using Eq. (S7)

$$\|\Omega\|_{\mathrm{HS}}^2 = \frac{1}{4}\sum_{\mu,\nu,\lambda,\kappa} f_{\mu\nu\lambda} f_{\mu\nu\kappa} b_\lambda b_\kappa = \frac{d}{2}\|b\|^2. \tag{S14}$$

Combining Eqs. (S11) and (S14) gives

$$\mathrm{tr}\, F + \frac{8}{d}\|\Omega\|_{\mathrm{HS}}^2 = 8\left(\frac{d^2-2}{d} + \langle \mathcal{S} \rangle\right). \tag{S15}$$

This is Eq. (3) of the main text. The Fisher-budget cap follows from $\langle \mathcal{S} \rangle \leq 1$ and $\|b\|^2 \geq 0$,

$$\mathrm{tr}\, F \leq \frac{8(d-1)(d+2)}{d}, \tag{S16}$$

Equality holds if and only if $\langle \mathcal{S} \rangle = 1$, $b = 0$.

The condition $\langle \mathcal{S} \rangle = 1$ is equivalent to $A = A^\top$, because $\mathcal{S}$ is an involution and the state is normalized. The condition $b = 0$ is equivalent to $\rho_a + \rho_b = \frac{2I_d}{d}$. Indeed, $\rho_a + \rho_b - \frac{2I_d}{d}$ is traceless Hermitian, and its expansion coefficients in the orthonormal basis $\{g_\mu\}$ are precisely $b_\mu$.

Define $\delta = \rho_a + \rho_b - \frac{2I_d}{d}$, $\Delta = \frac{8(d-1)(d+2)}{d} - \mathrm{tr}\, F$ and use $\|A - A^\top\|_{\mathrm{HS}}^2 = 2 - 2\mathbb{R}\mathrm{e}\,\mathrm{Tr}\,(A^\dagger A^\top) = 2 - 2\langle \mathcal{S} \rangle$, and $\|b\|^2 = \|\delta\|_{\mathrm{HS}}^2$. Thus, Eq. (S11) gives the deficit identity,

$$\Delta = 4\|A - A^\top\|_{\mathrm{HS}}^2 + 4\left\|\rho_a + \rho_b - \frac{2I_d}{d}\right\|_{\mathrm{HS}}^2. \tag{S17}$$

This is Eq. (4) of the main text.

**S3. Stabilizer kernel and Fisher-budget rigidity**

Let $h = \sum_\mu h_\mu g_\mu \in \mathfrak{su}(d)$ be a traceless Hermitian generator, and define $G_h = h \otimes I_d + I_d \otimes h$. The QFIM quadratic form is

$$h^\top F h = 4\mathrm{Var}_{\psi_A}(G_h). \tag{S18}$$

Therefore,

$$h \in \ker F \iff G_h|\psi_A\rangle = \alpha_h|\psi_A\rangle, \tag{S19}$$

where $\alpha_h = \langle G_h \rangle = \mathrm{Tr}[(\rho_a + \rho_b)h]$ is real. Using Eq. (S2), $G_h|\psi_A\rangle = |\psi_{hA + Ah^\top}\rangle$. Consequently, $h \in \ker F \Leftrightarrow hA + Ah^\top = \mathrm{Tr}[(\rho_a + \rho_b)h]A$. This is Eq. (1) of the main text and identifies $\ker F$ with the infinitesimal stabilizer Lie algebra of the probe ray under collective $SU(d)$ action.

Now impose Fisher-budget saturation. By Eq. (S17), this requires $A = A^\top$, $\rho_a + \rho_b = \frac{2I_d}{d}$. If $A = A^\top$, then $\rho_a = AA^\dagger$, $\rho_b = (A^\dagger A)^\top = AA^\dagger = \rho_a$. This gives $\rho_a = I_d/d$. By the Autonne-Takagi factorization [S1], every complex symmetric matrix can be written as $A = uDu^\top$, where $u \in U(d)$ and $D \geq 0$ is diagonal. The normalization and the condition $\rho_a = I_d/d$ force $D = \frac{I_d}{\sqrt{d}}$. Therefore the Fisher-budget maximizers are exactly

$$A = \frac{uu^\top}{\sqrt{d}},\ u \in U(d) \tag{S20}$$

This is Eq. (5) of the main text. Since all states in Eq. (S20) are related by collective $u \otimes u$ transformations, the QFIM rank is constant on this orbit. It is enough to evaluate the kernel at $A = \frac{I_d}{\sqrt{d}}$. At this representative,

$$G_h|\Phi\,\rangle = \frac{1}{\sqrt{d}}|\psi_{h+h^\top}\rangle, \tag{S21}$$

Thus, $h \in \ker F \Leftrightarrow h + h^\top = 0$. For Hermitian $h$, this means that $h$ is purely imaginary and antisymmetric. The real dimension of this space is $\frac{d(d-1)}{2}$. Hence every Fisher-budget maximizer has

$$F = \frac{16}{d}\Pi_{\mathfrak{p}}, \quad \operatorname{corank} F = \dim \mathfrak{so}(d) = \frac{d(d-1)}{2} \tag{S22}$$

The first identity in Eq. (S22) is Eq. (6) of the main text. The visible dimension is $\dim \mathfrak{p} = \frac{(d-1)(d+2)}{2}$ visible directions, each with Fisher eigenvalue $16/d$.

**S4. Exact Maximally entangled boundary spectrum and antiunitary classification**

A general maximally entangled bipartite probe can be written as

$$|\psi_U\rangle = (I_d \otimes U)|\Phi\rangle, \quad U \in U(d), \tag{S23}$$

and therefore, has coefficient matrix $A = \frac{U^\top}{\sqrt{d}}$,

**A. Stabilizer equation and antiunitary square**

Using Eq. (S3), the stabilizer condition Eq. (S19) reduces to

$$Uh^\top + hU = cU. \tag{S24}$$

Right-multiplying by $U^\dagger$ gives $Uh^\top U^\dagger + h = cI_d$. Taking the trace, $cd = \mathrm{Tr}(Uh^\top U^\dagger) + \mathrm{Tr}(h) = \mathrm{Tr}(h^\top) + \mathrm{Tr}(h) = 0$. Hence $c = 0$. Since $h^\top = \bar{h}$ for Hermitian $h$,

$$U\bar{h}U^\dagger = -h \tag{S25}$$

which is Eq. (7) of the main text. Define the antiunitary

$$J = UC, \quad V = J^2 = U\bar{U} \tag{S26}$$

where $C$ denotes computational-basis conjugation. Eq. (S25) becomes $JhJ^{-1} = -h$, therefore,

$$\ker F = \{h = h^\dagger : JhJ^{-1} = -h\} \tag{S27}$$

The tracelessness is automatic. Since $h = h^\dagger$ satisfies Eq. (S27), then $\overline{\mathrm{Tr}(h)} = \mathrm{Tr}(JhJ^{-1}) = -\mathrm{Tr}\, h$. $\mathrm{Tr}\, h = 0$, as it is real. For a maximally entangled probe, $b = 0$, and $\langle S\rangle = \frac{1}{d}\mathrm{Tr}(U\bar{U}) = \frac{1}{d}\mathrm{Tr}(V)$, substituting into Eq. (S11) gives

$$\mathrm{tr}\, F_U = \frac{8}{d}(d^2 - 2 + \mathrm{Tr}\, V). \tag{S28}$$

This is Eq. (8) of the main text.

**B. Complete boundary QFIM**

Let $\mathcal{H}_d = \{h = h^\dagger : \mathrm{Tr}\, h = 0\}$ be the real Hilbert space of traceless Hermitian matrices, equipped with Hilbert-Schmidt inner product $\langle h, k\rangle_{\mathrm{HS}} = \mathrm{Tr}\,(hk)$. Define the real-linear map

$$K_U : \mathcal{H}_d \to \mathcal{H}_d,\ K_U(h) = U^\top \bar{h}\bar{U}. \tag{S29}$$

The map $K_U$ is real orthogonal. Indeed, $\langle K_U(h), K_U(k)\rangle_{\mathrm{HS}} = \mathrm{Tr}\left(U^\top \bar{h}\bar{U}U^\top \bar{k}\bar{U}\right) = \mathrm{Tr}\left(\bar{h}\bar{k}\right) = \mathrm{Tr}\,(hk)$. The tangent coefficient matrix is

$$hA + Ah^\top = \frac{1}{\sqrt{d}}\left(hU^\top + U^\top \bar{h}\right) = \frac{1}{\sqrt{d}}(h + K_U h)U^\top. \tag{S30}$$

Because $b = 0$, the expectation value of every traceless collective generator vanishes. Therefore, $\langle h, F_U h\rangle_{\text{HS}} = 4\|hA + Ah^\top\|_{\text{HS}}^2 = \frac{4}{d}\|(I + K_U)h\|_{\text{HS}}^2$. Hence the complete boundary QFIM is

$$F_U = \frac{4}{d}(I + K_U)^*(I + K_U), \tag{S31}$$

Where $*$ denotes the adjoint on the real Hilbert space $\mathcal{H}_d$. Every real orthogonal operator decomposes into $\pm 1$ lines and two-dimensional rotation blocks. If $\theta_k \in [0, \pi]$ are the corresponding canonical angles, counted with real multiplicity, then $(I + K_U)^*(I + K_U) = 2(1 + \cos\theta_k)I$ on the associated block. Therefore,

$$\operatorname{spec} F_U = \left\{\frac{8}{d}(1 + \cos\theta_k)\right\}_{k=1}^{d^2-1}. \tag{S32}$$

Eqs. (S31) and (S32) constitute Eq. (9) of the main text.

The map $K_U$ also satisfies

$$K_U^2(h) = V^\dagger h V, \quad \operatorname{Tr}_{\mathbb{R}} K_U = \operatorname{Tr} V - 1. \tag{S33}$$

The first identity follows from $K_U^2(h) = U^\top U^\dagger h U \bar{U} = V^\dagger h V$. The second follows by evaluating the real trace of $K_U$ in an orthonormal Hermitian basis and applying Eq. (S5). Taking the trace of Eq. (S31) gives $\operatorname{tr} F_U = \frac{4}{d}[2(d^2 - 1) + 2\operatorname{Tr}_{\mathbb{R}} K_U]$, which reproduces Eq. (S28). If $U' = wUw^\top$, then $K_{U'}$ and $K_U$ are orthogonally similar on $\mathcal{H}_d$. The unitary-congruence class of $U$ is fixed by the unitary-similarity class of its cosquare, equivalently by the spectral class of $V = U\bar{U}$ [S2]. Consequently, the canonical angles $\theta_k$, and hence the entire QFIM spectrum, are class functions of $V$. The antiunitary frame determines the orientation of the Fisher eigenspaces in generator space, whereas the spectral class of $V$ determines their eigenvalues and multiplicities. Since $1 + \cos\theta_k \in [0,2]$,

$$0 \preccurlyeq F_U \preccurlyeq \frac{16}{d} I_{\mathcal{H}_d}. \tag{S34}$$

**C. Isotropic projector strata**

At $V = \pm I_d$, $K_U^2 = I_{\mathcal{H}_d}$. Therefore, every canonical angle is either 0 or $\pi$. The $V = -I_d$ stratum exists only for even $d$. In either extreme stratum,

$$F_U = \frac{16}{d}\Pi_U. \tag{S35}$$

where $\Pi_U$ is the Hilbert-Schmidt orthogonal projector onto the $\mathrm{K}_U = +1$ sector. This is the projector form stated immediately before Eq. (10) of the main text.

At the Fisher-budget maximizer $V = I_d$, the blind sector has dimension $d(d-1)/2$, and therefore

$$\operatorname{spec} F = \left\{ \left(\frac{16}{d}\right)^{\times (d-1)(d+2)/2}, \quad 0^{\times d(d-1)/2} \right\}. \tag{S36}$$

This is Eq. (10) of the main text. Hence, the maximizing QFIM is isotropic on its identifiable subspace.

**D. Antiunitary corank classification**

The blind sector is a Lie algebra in the Hermitian-generator convention. If $h, k \in \ker F$ and $[h,k]_{\mathfrak{g}} \equiv -i[h,k]$, antiunitarity gives $J[h,k]_{\mathfrak{g}} J^{-1} \equiv -i[h,k]_{\mathfrak{g}}$. Therefore, $[h,k]_{\mathfrak{g}} \in \ker F$.

Decompose $V$, into conjugate eigenvalue pairs $e^{\pm i\phi_j}$ of multiplicity $m_j$, and real eigenspaces $V = \pm 1$ of dimensions $n_+$ and $n_-$, with $n_-$ even. Counting Hermitian solutions of Eq. (S25) sector by sector via the threefold way [S3], $V = +1$ sector contributes $\frac{1}{2} n_+(n_+ - 1)$, the $V = -1$ sector contributes $\frac{1}{2} n_-(n_- + 1)$ and each conjugate pair contributes $m_j^2$. Therefore

$$\operatorname{corank} F = \sum_j m_j^2 + \frac{1}{2} n_+(n_+ - 1) + \frac{1}{2} n_-(n_- + 1), \tag{S37}$$

Equivalently, the kernel algebra is

$$\operatorname{Ker} F \simeq \mathfrak{so}(n_+) \oplus \bigoplus_j \mathfrak{u}(m_j) \oplus \mathfrak{sp}\left(\frac{n_-}{2}\right), \tag{S38}$$

The extreme cases are

$$V = I_d: \quad \operatorname{corank} F = \tfrac{d(d-1)}{2}; \quad V = -I_d,\, d \text{ even: } \operatorname{corank} F = \tfrac{d(d+1)}{2}, \tag{S39}$$

For generic $V$, all nonreal eigenvalues are simple conjugate pairs, and when $d$ is odd the lone real eigenvalue is $+1$, since $\det V = |\det U|^2 = 1$. Therefore,

$$\operatorname{corank} F \geq \left\lfloor \frac{d}{2} \right\rfloor, \tag{S40}$$

with equality on the generic maximally entangled boundary stratum.

To illustrate how the antiunitary class controls the complete boundary spectrum, consider the $d = 4$ maximally entangled family $U(\alpha) = R_{12}(\alpha) \oplus I_2$, $A(\alpha) = U(\alpha)^{\top}/2$, $0 \leq \alpha \leq \pi/2$. Since $U(\alpha)$ is real, $V(\alpha) = U(\alpha)\overline{U(\alpha)} = R_{12}(2\alpha) \oplus I_2$. The five spectral branches are $0^{\times 2}$, $[2(1 - \cos\alpha)]^{\times 4}$, $[2(1 + \cos 2\alpha)]^{\times 2}$, $[2(1 + \cos\alpha)]^{\times 4}$, $4^{\times 3}$. As shown in Fig. S1, the Fisher-budget maximizer at $\alpha = 0$ has corank 6; every interior point $0 < \alpha < \pi/2$ has the generic corank 2; and the endpoint $\alpha = \pi/2$, where $V = (-I_2) \oplus I_2$ , has corank 4. The crossing at $\alpha = \pi/3$ exchanges the ordering of two nonzero branches but does not change the boundary corank

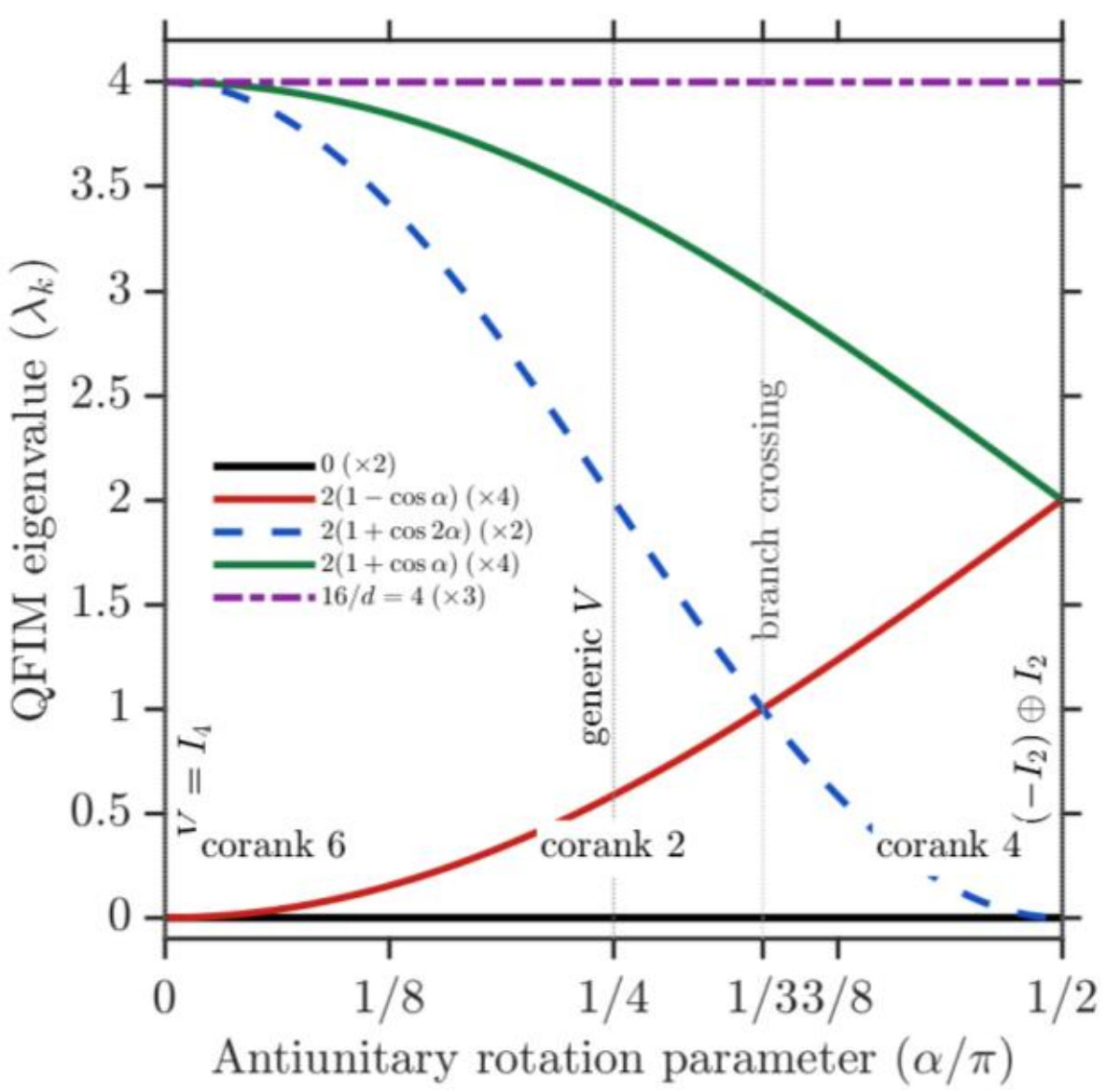

FIG. S1. Complete QFIM spectrum for $U(\alpha)$, $A(\alpha) = U(\alpha)^{\top}/2$ with $V(\alpha) = U(\alpha)\overline{U(\alpha)}$. The spectrum consists of the five branches listed in the text. At $\alpha = 0$, $V = I_4$ and the Fisher-budget maximizer has corank 6. For $0 < \alpha < \pi/2$, the family lies on the generic corank-2 stratum. At $\alpha = \pi/2$, $V = (-I_2) \oplus I_2$ and the corank is 4. The two nonzero branches cross at $\alpha = \pi/3$, while the upper branch remains fixed at the boundary ceiling $16/d = 4$.

Hence, no maximally entangled bipartite probe is fully identifiable under collective $SU(d)$ encoding.

**S5. Weak Compatibility forces singularity and quantitative near-blindness**

Let $\rho_a = AA^{\dagger}$, $\rho_b = (A^{\dagger}A)^{\top}$. For traceless Hermitian $h$, $\langle G_h \rangle = \mathrm{Tr}[(\rho_a + \rho_b)h]$. Define $\delta \coloneqq \rho_a + \rho_b - \frac{2I_d}{d}$. Since the coordinates of $\delta$ in the basis $\{g_\mu\}$ are $b_\mu$,

$$b = 0 \quad \Leftrightarrow \quad \delta = 0 \quad \Leftrightarrow \Omega = 0. \tag{S41}$$

Now define $h_0 = \rho_a - \frac{I_d}{d}$. Using $\rho_a A = AA^{\dagger}A = A(A^{\dagger}A) = A\bar{\rho}_b$, where $\bar{\rho}_b = A^{\dagger}A$, we obtain

$$h_0 A + A h_0^{\top} = A\overline{\delta}. \tag{S42}$$

Equivalently, $G(h_0)|\psi_A\rangle = |\psi_{A\overline{\delta}}\rangle$.

**A. Exact directional identity**

The norm of the tangent is $\left\|A\bar{\delta}\right\|_{\mathrm{HS}}^{2} = \mathrm{Tr}\left[\left(A\bar{\delta}\right)^{\dagger}A\bar{\delta}\right] = \mathrm{Tr}(\delta\rho_b\delta)$. Moreover, $\langle G_{h_0}\rangle = \mathrm{Tr}\left[\left(\frac{2I_d}{d} + \delta\right)h_0\right] = \mathrm{Tr}(\delta h_0)$, because $\mathrm{Tr}\,h_0 = 0$. Therefore,

$$h_0^{\top}Fh_0 = 4[\mathrm{Tr}(\delta\rho_b\delta) - (\mathrm{Tr}(\delta h_0)^2)]. \tag{S43}$$

If $\Omega = 0$, then $\delta = 0$, and Eq. (S42) gives $G_{h_0}\,|\psi_A\rangle = 0$. If $h_0 \neq 0$, this produces a nonzero element of $\ker F$. If $h_0 = 0$, then $\rho_a = \frac{I_d}{d}$ and $\delta = 0$ also gives $\rho_b = \frac{I_d}{d}$. The probe is maximally entangled and is singular by Sec. S4. Hence,

$$\Omega = 0 \implies \ker F \neq \{0\}. \tag{S44}$$

Eq. (S44) is the preliminary compatibility-singularity statement.

**B. Quantitative identifiability bound**

From Eq. (S43), $h_0^{\top}Fh_0 \leq 4\mathrm{Tr}(\delta\rho_b\delta) \leq 4\|\delta\|_{\mathrm{HS}}^2$. The first inequality drops the nonnegative squared-expectation subtraction in Eq. (S43). The last inequality uses $0 \preccurlyeq \rho_b \preccurlyeq I_d$. Since $\|\delta\|_{\mathrm{HS}}^2 = \|b\|^2 = \frac{2}{d}\|\Omega\|_{\mathrm{HS}}^2$, we obtain $h_0^{\top}Fh_0 \leq \frac{8}{d}\|\Omega\|_{\mathrm{HS}}^2$. By the variational characterization of $\lambda_{\min}(F)$, $\lambda_{\min}(F)\|h_0\|_{\mathrm{HS}}^2 \leq h_0^{\top}Fh_0$. Therefore, whenever $\rho_a \neq I_d/d$,

$$\lambda_{\min}(F)\left\|\rho_a - \frac{I_d}{d}\right\|_{\mathrm{HS}}^2 \leq \frac{8}{d}\|\Omega\|_{\mathrm{HS}}^2 \tag{S45}$$

This is Eq. (12) of the main text. Exchanging the two subsystems gives the analogous bound with $\rho_b$. A state-dependent refinement follows from $\mathrm{Tr}(\delta\rho_b\delta) = \mathrm{Tr}(\rho_b\delta^2) \leq \|\rho_b\|_{\infty}\|\delta\|_{\mathrm{HS}}^2$

$$\lambda_{\min}(F)\left\|\rho_a - \frac{I_d}{d}\right\|_{\mathrm{HS}}^2 \leq \frac{8}{d}\|\rho_b\|_{\infty}\|\Omega\|_{\mathrm{HS}}^2 \tag{S46}$$

For nonsingular $F$, the same direction gives a cost lower bound. Let $\hat{h}_0 = h_0/\|h_0\|_{\mathrm{HS}}$. Then $\mathrm{tr}\,F^{-1} \geq \hat{h}_0{}^{\top}F^{-1}\hat{h}_0 \geq \frac{1}{\left(\hat{h}_0{}^{\top}F\hat{h}_0\right)}$, therefore,

$$C^{\mathrm{H}}(I) \geq C^{\mathrm{SLD}} = \mathrm{tr}\,F^{-1} \geq \frac{d\,\left\|\rho_a - \frac{I_d}{d}\right\|_{\mathrm{HS}}^2}{8\|\Omega\|_{\mathrm{HS}}^2}, \tag{S47}$$

Eq. (S47) is Eq. (12a) of the main text. At fixed nonzero local mixedness defect, Eq. (S45) implies $\lambda_{\min}(F) = O(\|\Omega\|_{\mathrm{HS}}^2)$ as $\Omega \to 0$.

**C. Universal compatibility-induced corank floor**

Suppose $\Omega = 0$, so that $\rho_a + \rho_b = \frac{2\, I_d}{d}$. Apply a collective basis transformation so that $\rho_a$ is diagonal. The same transformation diagonalizes $\rho_b$, because $\rho_b = \frac{2\, I_d}{d} - \rho_a$. Let $E_\lambda$ denote the eigenspace of $\rho_a$ with eigenvalue $\lambda$. Since $AA^\dagger = \rho_a$, $A^\dagger A = \bar{\rho}_b$, the matrix $A$ maps $E_{2/d-\lambda}$ onto $E_\lambda$. The multiplicities of $\lambda$ and $2/d - \lambda$ are therefore equal. If $h \in \ker F$, compatibility reduces the stabilizer equation to $hA + Ah^\top = 0$. This equation and its adjoint imply $[h, \rho_a] = 0$. Hence, $h$ is block diagonal in the eigenspace decomposition of $\rho_a$.

Consider a non-self-paired eigenvalue pair $\lambda^\# = 2/d - \lambda \neq \lambda$, with common multiplicity $m$. For $0 < \lambda < 2/d$, write the corresponding blocks as $A_{\lambda,\lambda^\#} = \sqrt{\lambda} U_+$ and $A_{\lambda^\#,\lambda} = \sqrt{\lambda^\#} U_-$, where $U_\pm \in U(m)$. If $X$ and $Y$ are the Hermitian blocks of $h$ on $E_\lambda$ and $E_{\lambda^\#}$, the stabilizer equation gives $XU_+ + U_+ Y^\top = 0$ and $YU_- + U_- X^\top = 0$. . Eliminating $Y$ yields

$$[X, W_\lambda] = 0, \quad W_\lambda = U_-^\top U_+^\dagger \tag{S48}$$

If the eigenvalues of the unitary $W_\lambda$ have multiplicities $q_r$, the real dimension of its Hermitian commutant is $\sum_r q_r^2 \geq m$. Hence, every non-self-paired spectral pair of multiplicity $m$ contributes at least $m$ linearly independent null generators. Their traces cancel pairwise because the first stabilizer equation gives $\mathrm{Tr}\, Y = -\mathrm{Tr}\, X$. The endpoint pair $\lambda = 0$, $\lambda^\# = 2/d$ contributes $m^2 \geq m$ null generators by the same block equation with one vanishing $A$-block. These endpoint solutions also satisfy $\mathrm{Tr}\, Y = -\mathrm{Tr}\, X$. Therefore, they are traceless on the paired sector.

The self-paired eigenspace $\lambda = 1/d$, of dimension $r$, satisfies $A_{1/d,1/d} = U_0/\sqrt{d}$ with $U_0 \in U(r)$. Its stabilizer equation is precisely the maximally entangled boundary problem in dimension $r$, and Sec. S4 gives at least $\lfloor r/2 \rfloor$ independent null generators. Writing the total dimension as $d = 2\sum_{\lambda<1/d} m_\lambda + r$, we obtain

$$\operatorname{corank} F \geq \sum_{\lambda<1/d} m_\lambda + \left\lfloor \frac{r}{2} \right\rfloor = \left\lfloor \frac{d}{2} \right\rfloor. \tag{S49}$$

Eq. (S49) is Eq. (11) of the main text. The bound is sharp. For even $d$, choose $d/2$ one-dimensional complementary spectral pairs with distinct complementary eigenvalues and generic phases in each paired block. For odd $d$, use $(d-1)/2$ such pairs and one one-dimensional self-paired $1/d$ block.

**S6. Exchange-symmetric Holevo bound**

To derive the exchange-symmetric Fisher blocks and the closed-form Holevo bound used in Sec. V of the main text, reduce the exchange-symmetric probe by Autonne-Takagi factorization [S1] to

$$A = \mathrm{diag}(s_1, \dots, s_d),\ \ s_i = \sqrt{\lambda_i},\ \ \lambda_i > 0, \quad \textstyle\sum_i \lambda_i = 1. \tag{S50}$$

For $i < j$, define the normalized Hermitian off-diagonal generator $X_{ij} = \frac{|i\rangle\langle j| + |j\rangle\langle i|}{\sqrt{2}}$, $Y_{ij} = \frac{-i|i\rangle\langle j| + i|j\rangle\langle i|}{\sqrt{2}}$. Also define the normalized two-site symmetric vector

$$\left|\Sigma_{ij}\right\rangle = \frac{|ij\rangle + |ji\rangle}{\sqrt{2}}. \tag{S51}$$

Here $X_{ij}$ and $Y_{ij}$ are the symmetric and antisymmetric Hermitian off-diagonal generators, and $\left|\Sigma_{ij}\right\rangle$ is the corresponding normalized exchange-symmetric two-site vector. A direct computation gives

$$G_{X_{ij}}|\psi_A\rangle = (s_i + s_j)\left|\Sigma_{ij}\right\rangle, \quad G_{Y_{ij}}|\psi_A\rangle = i(s_i - s_j)\left|\Sigma_{ij}\right\rangle. \tag{S52}$$

Since $\left\langle\psi\middle|\Sigma_{ij}\right\rangle = 0$, the block is diagonal

$$F_{ij} = 4\begin{pmatrix} (s_i + s_j)^2 & 0 \\ 0 & (s_i - s_j)^2 \end{pmatrix}, \quad \left|\Omega_{ij}\right| = \left|\lambda_i - \lambda_j\right|. \tag{S53}$$

The off-diagonal blocks are mutually orthogonal, and they are orthogonal to the diagonal traceless, or Cartan, sector. Because the quantum geometric tensor and the identity cost are block diagonal in this decomposition, the pure-state Holevo variational problem [S4,S5] separates into independent root and Cartan sectors.

For one root block, write, $a = (s_i + s_j)$, $b = (s_i - s_j)$, $|\Sigma\rangle = \left|\Sigma_{ij}\right\rangle$. The tangent vectors $-ia|\Sigma\rangle$ and $b|\Sigma\rangle$ are orthogonal to $|\psi_A\rangle\rangle$, so the SLDs are

$$L_X = 2(-ia|\Sigma\rangle\langle\psi_A| + ia|\psi_A\rangle\langle\Sigma|), \qquad L_Y = 2b(|\Sigma\rangle\langle\psi_A| + |\psi_A\rangle\langle\Sigma|). \tag{S54}$$

Then $\langle[L_X, L_Y]\rangle = 8iab$, therefore

$$D_{XY} \coloneqq \frac{i}{2}\langle[L_X, L_Y]\rangle = -4ab, \quad \det F_{ij} = 16(\lambda_i - \lambda_j)^2. \tag{S55}$$

Hence $|D_{XY}| = 4\left|\lambda_i - \lambda_j\right| = \sqrt{\det F_{ij}}$. Each block saturates the normalized incompatibility bound, giving

$$C_{ij}^{\mathrm{H}} = \mathrm{tr}\, F_{ij}^{-1} + \frac{2}{\sqrt{\det F_{ij}}} = \mathrm{tr}\, F_{ij}^{-1} + \frac{1}{(2|\lambda_i - \lambda_j|)}\,. \tag{S56}$$

For a diagonal traceless generator $h = \mathrm{diag}(h_1, \dots, h_d)$, $G_h|\psi_A\rangle = \sum_k 2h_k\sqrt{\lambda_k}\,|kk\rangle$, and hence on traceless diagonal subspace,

$$F_C = 16(\mathrm{diag}\,\lambda - \lambda\lambda^\top),\ \mathrm{tr}\,F_C^{-1} = \frac{d-1}{16d}\sum_i \frac{1}{\lambda_i}. \tag{S57}$$

The second identity gives the Cartan contribution and is discussed in Sec. V of the main text. For $d = 2$, it reduces to $\mathrm{tr}\,F_C^{-1} = \frac{1}{32\lambda_1\lambda_2}$. Summing the off-diagonal Holevo blocks and the Cartan contribution gives the closed-form identity-cost Holevo bound,

$$C_{\mathrm{sym}}^{\mathrm{H}}(\lambda) = \sum_{i<j}\left[\frac{1}{4(s_i+s_j)^2} + \frac{1}{4(s_i-s_j)^2} + \frac{1}{(2|\lambda_i-\lambda_j|)}\right] + \frac{d-1}{16d}\sum_i \frac{1}{\lambda_i}. \tag{S58}$$

Subtracting $C_{\mathrm{sym}}^{\mathrm{SLD}} = \mathrm{tr}\,F^{-1}$gives

$$C_{\mathrm{sym}}^{\mathrm{H}} - C_{\mathrm{sym}}^{\mathrm{SLD}} = \sum_{i<j}\frac{1}{2|\lambda_i-\lambda_j|}. \tag{S59}$$

The $1/|\lambda_i-\lambda_j|$ incompatibility term is subleading compared with the $1/(s_i-s_j)^2$ SLD divergence near Schmidt degeneracy, but it is still singular. For $d = 2$, Eq. (S58) reproduces the known two-qubit collective result [S6].

**S7. Deficit-controlled divergence near the Fisher-budget cap**

Let $s_i = \sqrt{\lambda_i}$, $\sum_i s_i^2 = 1$. Define the Schmidt-overlap deficit $\Delta_* = d - \left(\sum_i \sqrt{\lambda_i}\right)^2$. The elementary identity $\sum_{i<j}(s_i-s_j)^2 = d\sum_i s_i^2 - (\sum_i s_i)^2$ gives

$$\Delta_* = \sum_{i<j}(s_i-s_j)^2. \tag{S60}$$

The exchange-symmetric SLD cost contains the soft antisymmetric-root contribution

$$C_{\mathrm{sym}}^{\mathrm{SLD}} \ge \sum_{i<j}\frac{1}{4(s_i-s_j)^2}. \tag{S61}$$

Let $M = \frac{d(d-1)}{2}$. By Cauchy-Schwarz,

$$\left(\sum_{i<j}\frac{1}{4(s_i-s_j)^2}\right)\Delta_* \ge \frac{M^2}{4}. \tag{S62}$$

Therefore,

$$\mathrm{tr}\,F^{-1} \ge \frac{M^2}{4\Delta_*} = \frac{\left[\frac{d(d-1)}{2}\right]^2}{4\Delta_*}. \tag{S63}$$

Eq. (S63) gives the first lower bound in Eq. (13) of the main text. Near the Fisher-budget cap on the exchange-symmetric manifold, write $\lambda_i = \frac{1}{d} + x_i$, $\sum_i x_i = 0$, $\sum_i x_i^2 \to 0$. Then

$$s_i = \frac{1}{\sqrt{d}} + \frac{\sqrt{d}}{2} x_i - \frac{d^{\frac{3}{2}}}{8} x_i^2 + O\left(d^{\frac{5}{2}} |x_i|^3\right). \tag{S64}$$

Hence

$$\sum_i s_i = \sqrt{d} - \frac{d^{\frac{3}{2}}}{8} \sum_i x_i^2 + o\left(\sum_i x_i^2\right), \tag{S65}$$

and

$$\Delta_* = \frac{d^2}{4} \sum_i x_i^2 + o\left(\sum_i x_i^2\right). \tag{S66}$$

For exchange-symmetric probes,

$$\Delta = \frac{8(d-1)(d+2)}{d} - \operatorname{tr} F = 16 \sum_i \left(\lambda_i - \frac{1}{d}\right)^2 = 16 \sum_i x_i^2. \tag{S67}$$

Therefore,

$$\Delta_* = \frac{d^2}{64} \Delta[1 + o(1)], \tag{S68}$$

Eq. (S68) is the asymptotic relation used to obtain the second lower bound in main-text Eq. (13). Combining Eqs. (S63) and (S68) gives

$$\operatorname{tr} F^{-1} \geq \frac{4(d-1)^2}{\Delta} [1 + o(1)]. \tag{S69}$$

Eq. (S63) and (S69), together with $C^{\mathrm{H}}(I) \geq C^{\mathrm{SLD}} = \operatorname{tr} F^{-1}$, constitute Eq. (13) of the main text. Hence, Holevo cost also diverges inversely in the Fisher-budget deficit.

**S8. Optimal soft-mode opening near the Fisher-budget cap**

Consider an exchange-symmetric path $\lambda_i(\varepsilon) = \frac{1}{d} + \varepsilon x_i$, $\sum_i x_i = 0$, with pairwise distinct $x_i$. Expanding Eq. (S53),

$$4(s_i - s_j)^2 = d(x_i - x_j)^2 \varepsilon^2 + O(\varepsilon^3). \tag{S70}$$

Let $g = \min_{i<j} |x_i - x_j|$. The weakest Fisher eigenvalue therefore satisfies $\lambda_{\min}(F) \leq d g^2 \varepsilon^2 + O(\varepsilon^3)$. For $d$ ordered real numbers with minimum spacing $g$ and zero mean,

$$\sum_i x_i^2 \geq \frac{d(d^2-1)}{12} g^2, \tag{S71}$$

with equality for an equally spaced sequence. Using $\Delta = 16\,\varepsilon^2 \sum_i x_i^2$, we obtain

$$\lambda_{\min}(F) \leq \frac{3}{(4(d^2-1))}\,\Delta[1+o(1)]. \tag{S72}$$

The equally spaced spectral direction saturates the coefficient. Thus, even the stiffest exchange-symmetric approach to the Fisher-budget cap leaves the weakest Fisher eigenvalue at most linear in the budget deficit.

**S9. Probe-state diagnostics**

The quantities entering the exact budget identities and the quantitative near-blindness bound are properties of the input probe rather than of the encoded output family. The reduced density matrices follow from one-body tomography,

$$\rho_a = AA^\dagger, \rho_\mathrm{b} = (A^\dagger A)^\top. \tag{S73}$$

The collective polarization is then $b_\mu = \mathrm{Tr}\big[(\rho_a + \rho_b) g_\mu\big]$, and the weak-incompatibility norm follows from

$$\|\Omega\|_{HS}^2 = \frac{\mathrm{d}}{2}\|\mathrm{b}\|^2. \tag{S74}$$

The exchange expectation

$$\langle \mathcal{S} \rangle = Tr\,(A^\dagger A^\top), \tag{S75}$$

is the expectation value of the subsystem-swap observable and can be obtained through a direct swap measurement, a controlled-SWAP protocol, or an equivalent two-particle interference measurement, depending on the platform. Therefore, Eq. (S11) determines the complete Fisher budget from $\langle \mathcal{S} \rangle$ and $b$; Eq. (S17) determines the budget deficit from exchange asymmetry and the reduced-state sum; and Eq. (S45) bounds the smallest Fisher eigenvalue using $\rho_a$ and $\Omega$. No process tomography of the unknown collective transformation is required for these probe diagnostics.

**S10. Numerical construction and verification details**

All numerical data in Figs. 1 and 2 (main text) were generated directly from the pure-state quantum geometric tensor

$$Q_{\mu\nu} = \langle \psi_A | \Delta G_\mu \Delta G_\nu | \psi_A \rangle,\; F = 4\mathbb{R}\mathrm{e}Q,\; \Omega = \mathbb{I}\mathrm{m}Q. \tag{S76}$$

The traceless Hermitian generators were constructed as the generalized Gell-Mann basis normalized according to Eq. (S4).

For Fig. 1(a) (main text), probes were sampled from normalized complex Ginibre matrices, $A = X/\sqrt{\mathrm{Tr}(XX^\dagger)}$, together with exchange-symmetric samples and perturbations of the maximizing orbit. The compatible nonmaximal qutrit marker was generated from

$$A_{\mathrm{comp}} = \mathrm{diag}\left(\sqrt{\frac{1}{3}+t}, \sqrt{\frac{1}{3}}, \sqrt{\frac{1}{3}-t}\right)P,\ P = \begin{pmatrix} 0 & 0 & 1 \\ 0 & 1 & 0 \\ 1 & 0 & 0 \end{pmatrix}, \tag{S77}$$

with $t = 0.17$. It satisfies

$$\rho_a + \rho_b = \frac{2I_3}{3}, \tag{S78}$$

and therefore $\Omega = 0$, while its QFIM is singular and nonmaximal.

For Fig. 1(b), Haar-random unitary matrices were obtained by QR decomposition of complex Ginibre matrices, with the phases of the diagonal entries of the triangular factor removed. Each maximally entangled probe was represented by

$$A = \frac{U^\top}{\sqrt{d}}. \tag{S79}$$

The theoretical markers were computed independently from

$$F_U^{\mathrm{th}} = \frac{4}{d}(I + K_U)^*(I + K_U), \tag{S80}$$

in the real Gell-Mann basis.

The implementation verifies the following identities numerically:

1. Eqs. (S11), (S14), (S15), and (S17);
2. the maximizing projector spectrum Eq. (S36);
3. the complete boundary spectrum Eq. (S32);
4. invariance of the boundary spectrum under $U \mapsto wUw^\top$;
5. Eqs. (S43) and (S45), and the universal corank floor Eq. (S49);
6. the exchange-symmetric block formulas Eqs. (S53) and (S56);
7. the qutrit coefficients derived in Sec. S11;
8. the multipartite QFIM Eq. (S104).

The deterministic identities are reproduced to floating-point errors of order $10^{-14} - 10^{-15}$. Eigenvalues whose absolute values are below $10^{-8}$ times the largest Fisher eigenvalue are treated as numerical zeros when reporting coranks.

Moreover, Fig. S1 was generated independently from the complete pure-state quantum geometric tensor and compared pointwise with the canonical-angle spectrum in Eq. (S32). Fig. S2 was obtained by direct diagonalization of the complete qutrit QFIM along the two fixed-spectrum paths in Sec. S11.

## S11. Codimension-controlled divergence and the qutrit fixed-spectrum example

### A. Symmetric-path amplitude formula

Consider an exchange-symmetric path $\lambda_i(\varepsilon) = \frac{1}{d} + \varepsilon x_i$, $\sum_i x_i = 0$, with pairwise distinct $x_i$. From Eq. (S70), the soft antisymmetric eigenvalues are $d\left(x_i - x_j\right)^2 \varepsilon^2 + O(\varepsilon^3)$. Hence,

$$\varepsilon^2 \operatorname{tr} F^{-1} \to \mathcal{A}(x), \ \mathcal{A}(x) = \frac{1}{d} \sum_{i<j} \frac{1}{\left(x_i - x_j\right)^2}. \tag{S81}$$

This gives the leading SLD divergence amplitude for an arbitrary regular exchange-symmetric approach direction.

### B. General codimension-controlled opening

Le $K = \ker F(A_0)$, $R = K^{\perp}$, $\kappa_V = \dim K$, with $A_0 = A_\gamma(0)$ maximally entangled of antiunitary class $V$. Here $K$ is the soft Fisher kernel at the maximally entangled boundary point, $R$ is its Hilbert-Schmidt orthogonal complement, and $\kappa_V$ is the boundary corank fixed by the antiunitary class. In the splitting $K \oplus R$,

$$F(\varepsilon) = \begin{pmatrix} \varepsilon^2 F_{KK}^{(2)} + \mathcal{O}(\varepsilon^3) & \varepsilon F_{KR}^{(1)} + \mathcal{O}(\varepsilon^2) \\ \varepsilon F_{RK}^{(1)} + \mathcal{O}(\varepsilon^2) & F_{RR}^{(0)} + \mathcal{O}(\varepsilon) \end{pmatrix} \tag{S82}$$

The $KK$ block is $\mathcal{O}(\varepsilon^2)$ because each null variance is minimized at $\varepsilon = 0$, the $KR$ block is $\mathcal{O}(\varepsilon)$, and $F_{RR}^{(0)} > 0$. The effective soft block is the Schur complement

$$B_\gamma = F_{KK}^{(2)} - F_{KR}^{(1)} \left(F_{RR}^{(0)}\right)^{-1} F_{RK}^{(1)}. \tag{S83}$$

Here $B_\gamma$ is the second-order effective Fisher matrix on the soft kernel $K$. For a regular path, meaning $B_\gamma > 0$, exactly $\kappa_V$ eigenvalues open as $\lambda_k(F) = c_k \varepsilon^2 + \mathcal{O}(\varepsilon^3)$, $c_k > 0$, with $c_k$ the eigenvalues of $B_\gamma$. Hence

$$\operatorname{tr}\left[\ F(A_\gamma(\varepsilon))^{-1}\right] = \frac{\alpha_\gamma}{\varepsilon^2} + \mathcal{O}(\varepsilon^{-1}), \ \alpha_\gamma = \sum_{k=1}^{\kappa_V} \frac{1}{c_k} \tag{S84}$$

The number $\kappa_V$ is class-invariant, while $\alpha_\gamma$ is path-dependent. For the qutrit fixed-spectrum example, take

$$D_\varepsilon = \mathrm{diag}\left(\sqrt{\tfrac{1}{3}+\varepsilon}, \sqrt{\tfrac{1}{3}}, \sqrt{\tfrac{1}{3}-\varepsilon}\right),\ 0<\varepsilon<\tfrac{1}{3}. \tag{S85}$$

For the exchange-symmetric path $A_{\mathrm{sym}}(\varepsilon) = D_\varepsilon \to V = I_3$, the three soft Fisher coefficients are $(3,3,12)$. Direct expansion of the exact formula Eq. (S58) gives

$$C^{\mathrm{H}}_{\mathrm{sym}}(\varepsilon) = \frac{3}{4\varepsilon^2} + \mathcal{O}(\varepsilon^{-1}). \tag{S86}$$

For the reoriented path $A_{\mathrm{rot}}(\varepsilon) = D_\varepsilon R_{13}(\pi/4)$, where

$$R_{13}(\alpha) = \begin{pmatrix} \cos\alpha & 0 & \sin\alpha \\ 0 & 1 & 0 \\ -\sin\alpha & 0 & \cos\alpha \end{pmatrix}, \tag{S87}$$

the Schmidt spectrum is unchanged, $A_{\mathrm{rot}}A_{\mathrm{rot}}^\dagger = D_\varepsilon^2$. At $\varepsilon = 0$, the associated unitary in the convention $|\psi_U\rangle = (I_d \otimes U)|\Phi\rangle$ is $U = R_{13}(-\pi/4)$, hence

$$V = U\overline{U} = R_{13}\left(-\frac{\pi}{2}\right). \tag{S88}$$

Here $V = U\overline{U}$ is the antiunitary-square invariant in this convention. With eigenvalues $\{i, -i, 1\}$, one conjugate pair and $n_+ = 1$. Therefore, Eq. (S37) gives corank one. Second-order perturbation theory for the simple zero eigenvalue of $F(0)$ gives

$$\lambda_{\min}(F) = 12\varepsilon^2 + \mathcal{O}(\varepsilon^3). \tag{S89}$$

A direct expansion of the pure-state Holevo program shows that the remaining inverse-Fisher terms are $O(1)$ and the Holevo surcharge is $O(\varepsilon^{-1})$. Therefore,

$$C^{\mathrm{H}}_{\mathrm{rot}}(\varepsilon) = \tfrac{1}{12\varepsilon^2} + \mathcal{O}(\varepsilon^{-1}),\ \tfrac{C^{\mathrm{H}}_{\mathrm{sym}}}{C^{\mathrm{H}}_{\mathrm{rot}}} \to 9, \tag{S90}$$

This gives the ninefold qutrit comparison discussed in the main text.

Fig. S2 resolves the ninefold cost ratio in Eq. (S90) into its individual soft Fisher channels. For the exchange-symmetric path $A_{\mathrm{sym}}(\varepsilon) = D_\varepsilon$, three eigenvalues vanish quadratically, $\lambda_{1,2}(F) = 3\varepsilon^2 + O(\varepsilon^3)$ and $\lambda_3(F) = 12\varepsilon^2 + O(\varepsilon^3)$. For the reoriented path $A_{\mathrm{rot}}(\varepsilon) = D_\varepsilon R_{13}(\pi/4)$, only one Fisher eigenvalue is soft, $\lambda_1(F) = 12\varepsilon^2 + O(\varepsilon^3)$. Therefore, the symmetric path receives three leading inverse-eigenvalue contributions, $\frac{1}{3}+\frac{1}{3}+\frac{1}{12}=\frac{3}{4}$, whereas the rotated path receives only $1/12$. The ninefold difference therefore originates from the number of soft identifiable channels fixed by the limiting antiunitary class, rather than from the Schmidt spectrum, which is identical along the two paths.

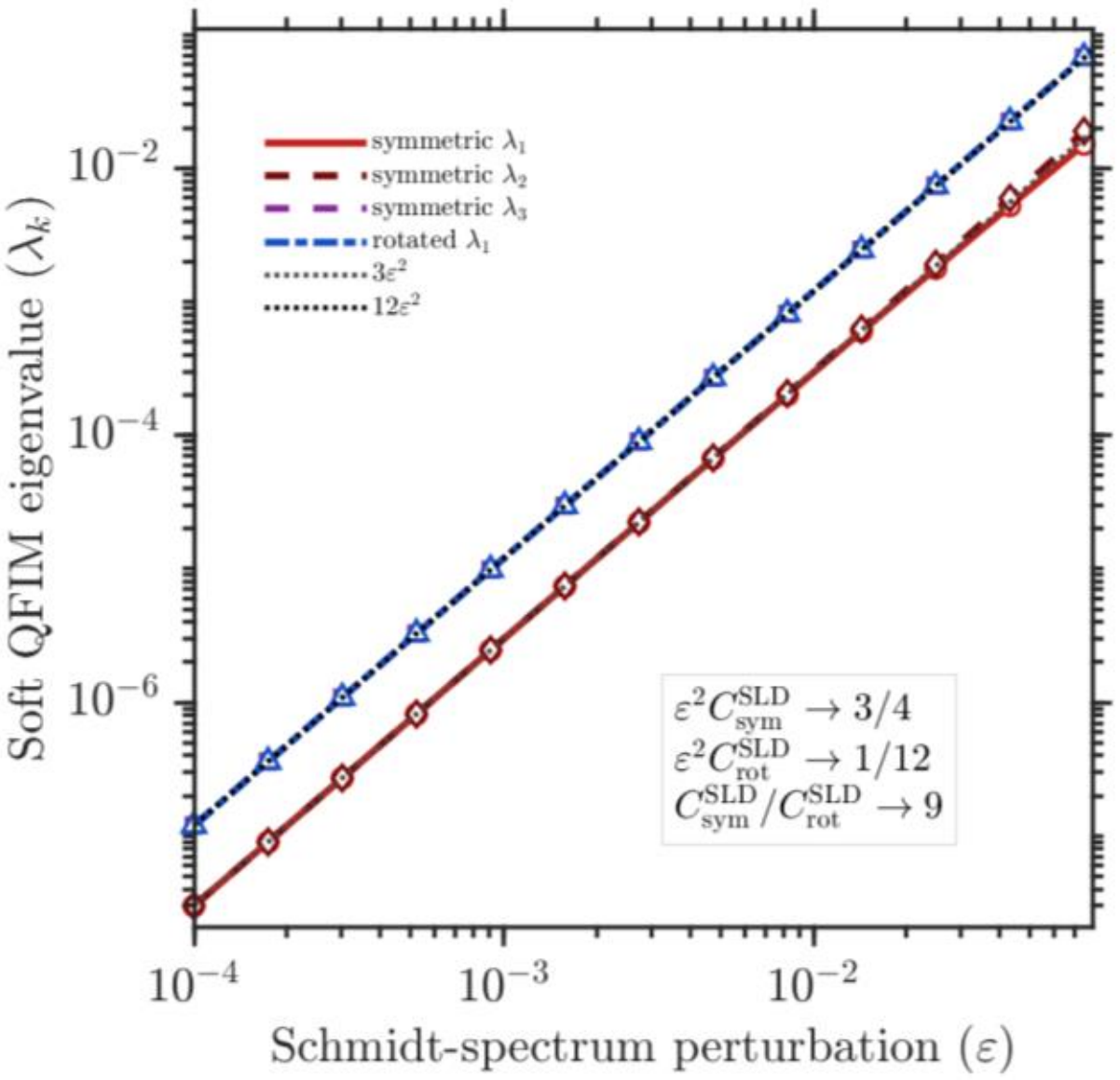

FIG. S2. Soft QFIM eigenvalues for two probe paths with the same Schmidt spectrum $\left(\frac{1}{3}+\varepsilon, \frac{1}{3}, \frac{1}{3}-\varepsilon\right)$. For the exchange-symmetric path $A_{\mathrm{sym}}(\varepsilon)$, the two lowest eigenvalues are degenerate and satisfy $\lambda_{1,2}(F) = 3\varepsilon^2 + O(\varepsilon^3)$., while the third satisfies $\lambda_3(F) = 12\varepsilon^2 + O(\varepsilon^3)$. For the reoriented path $A_{\mathrm{rot}}(\varepsilon)$ only one eigenvalue is soft, with $\lambda_1(F) = 12\varepsilon^2 + O(\varepsilon^3)$. The dotted lines show the asymptotic laws $3\varepsilon^2$ and $12\varepsilon^2$. Consequently, $\varepsilon^2 C_{\mathrm{sym}}^{\mathrm{SLD}} \to 3/4$, $\varepsilon^2 C_{\mathrm{rot}}^{\mathrm{SLD}} \to 1/12$, and the leading cost ratio tends to 9.

The same expansion gives the coefficient 12 for the family $R_{13}(\alpha)$, $\alpha \in (0, \pi/2)$. More generally, antiunitary class fixes the number of soft modes, while amplitudes are path-dependent.

The same leading coefficients govern both the SLD and Holevo divergences, because the Holevo surcharge is subleading at order $O(\varepsilon^{-1})$; Fig. 2(b) of the main text displays the corresponding SLD-cost separation; the Holevo cost share the same leading coefficients. Hence, two probes with the same Schmidt spectrum and the same entanglement can differ ninefold in their leading attainable cost because they approach different antiunitary boundary classes.

**S12. Multipartite budget law and sharpness of the bipartite obstruction**

We record the $N$-system extension only to delimit the scope of the bipartite obstruction in collective quantum metrology [S7]. Let

$$|\Psi\rangle \in (\mathbb{C}^d)^{\otimes N}, \quad G_\mu^{(N)} = \sum_{r=1}^{N} g_\mu^{(r)}, \qquad b_\mu = \langle G_\mu^{(N)} \rangle. \tag{S91}$$

Here $g_\mu^{(r)}$ denotes the generator $g_\mu$ acting on tensor factor $r$, and $b = (b_\mu)$ is the collective Bloch vector. For a pure unitary model,

$$\operatorname{tr} F = 4 \sum_\mu \left( \langle \left( G_\mu^{(N)} \right)^2 \rangle - b_\mu^2 \right). \tag{S92}$$

Using Eq. (S6)

$$\sum_\mu \left( G_\mu^{(N)} \right)^2 = \frac{N(d^2-1)}{d} I + 2 \sum_{1 \le r < s \le N} \left( S_{rs} - \frac{1}{d} I \right), \tag{S93}$$

where $S_{rs}$ swaps tensor factors $r$ and $s$. Hence

$$\operatorname{tr} F = \frac{4N(d^2-1)}{d} + 8 \sum_{r<s} \left( \langle S_{rs} \rangle - \frac{1}{d} \right) - 4\|b\|^2. \tag{S94}$$

The curvature relation has the same form as in the bipartite case

$$\Omega_{\mu\nu} = \tfrac{1}{2} \textstyle\sum_\lambda f_{\mu\nu\lambda} b_\lambda, \quad \|\Omega\|_{\mathrm{HS}}^2 = \tfrac{d}{2} \|b\|^2. \tag{S95}$$

Combining Eqs. (S94) and (S95) gives

$$\operatorname{tr} F + \frac{8}{d} \|\Omega\|_{\mathrm{HS}}^2 = \frac{4N(d^2-1)}{d} + 8 \sum_{r<s} \left( \langle S_{rs} \rangle - \frac{1}{d} \right) \tag{S96}$$

Since $\langle S_{rs} \rangle \le 1$,

$$\operatorname{tr} F + \frac{8}{d} \|\Omega\|_{\mathrm{HS}}^2 \le \frac{4N(d-1)(d+N)}{d}. \tag{S97}$$

Equality holds iff the probe is permutation-symmetric. Therefore,

$$\operatorname{tr} F \le \frac{4N(d-1)(d+N)}{d}, \tag{S98}$$

with equality iff $|\Psi\rangle \in \mathrm{Sym}^N(\mathbb{C}^d)$ and $b = 0$. For a permutation-symmetric pure state, this is equivalent to the one-body marginal condition

$$\rho_1 = \frac{I_d}{d}. \tag{S99}$$

If the probe is fully identifiable, then $F > 0$ on the $(d^2-1)$-dimensional parameter space. The arithmetic-harmonic mean inequality and Eq. (S98) give

$$\operatorname{tr} F^{-1} \geq \frac{(d^2-1)^2}{\operatorname{tr} F} \geq \frac{d(d-1)(d+1)^2}{4N(d+N)}. \tag{S100}$$

This is the $N$-system analogue of the bipartite SLD floor.

### A. Exact generalized-GHZ QFIM

For $N \geq 3$, consider

$$|\mathrm{GHZ}_{d,N}\rangle = \frac{1}{\sqrt{d}} \sum_{a=1}^{d} |\mathrm{a}\rangle^{\otimes N}. \tag{S101}$$

This is the first part of Eq. (14) of the main text. The state is permutation symmetric and has $\rho_1 = \frac{I_d}{d}$, $b = 0$. It therefore saturates Eq. (S98) and is weakly compatible. Decompose $\mathfrak{su}(d)$ into the $d(d-1)$ normalized off-diagonal generators and the $d-1$ Cartan generators. For an off-diagonal normalized generator, the collective action produces one-deviant-slot vectors. For distinct unordered level pairs these supports are orthogonal; within each pair $X/Y$, the real QFIM cross term vanishes. When $N \geq 3$, vectors associated with different base states, deviant states, slots, or generator pairs are mutually orthogonal. Their total squared norm is $N/d$, and their expectation value vanishes. Hence every normalized off-diagonal direction has QFI

$$F_{\mathrm{off}} = \frac{4N}{d}. \tag{S102}$$

For a normalized diagonal traceless generator
$h = \mathrm{diag}(h_1, \dots, h_d)$, $\sum_a h_a = 0$, $\sum_a h_a^2 = 1$, one has

$$G_h^{(N)} |\mathrm{GHZ}_{d,N}\rangle = \frac{N}{\sqrt{d}} \sum_a h_a |a\rangle^{\otimes N} \tag{S103}$$

Its expectation value vanishes and its squared norm is $N^2/d$. Therefore, every normalized Cartan direction has QFI $4N^2/d$. Cross terms vanish. Hence,

$$F_{\mathrm{GHZ}_{d,N}} = \frac{4N}{d} I_{\mathrm{off}} \oplus \frac{4N^2}{d} I_C, \quad N \geq 3. \tag{S104}$$

This is the second part of Eq. (14) of the main text. Here, $I_{\mathrm{off}}$ acts on the $d(d-1)$ off-diagonal generators and $I_{\mathrm{C}}$ acts on the $d-1$ Cartan generators. Hence,

$$\dim I_{\mathrm{off}} = d(d-1), \dim I_{\mathrm{C}} = d-1. \tag{S105}$$

Consequently,

$$\lambda_{\min}\left(F_{\mathrm{GHZ}_{d,N}}\right) = \frac{4N}{d} > 0, \ \operatorname{tr} F_{\mathrm{GHZ}_{d,N}} = \frac{4N(d-1)(d+N)}{d}. \tag{S106}$$

The full-rank property can also be seen directly from the stabilizer. Suppose

$$G_h^{(N)}\left|\mathrm{GHZ}_{d,N}\right\rangle = \alpha\left|\mathrm{GHZ}_{d,N}\right\rangle,\ h = h^\dagger,\ \mathrm{Tr}\, h = 0. \tag{S107}$$

For $\mathrm{N} \geq 3$, every off-diagonal matrix element $h_{ba}$, $b \neq a$ produces one-deviant-slot vectors outside the GHZ support. These vectors are mutually orthogonal, so Eq. (S107) forces $h_{ba} = 0$ ($b \neq a$). Therefore, $h$ is diagonal. The diagonal part gives $Nh_{aa} = \alpha$ for every $a$, so $h$ is proportional to $I_d$. Tracelessness gives $h = 0$. For $N = 1$, the pure-state orbit lies in $\mathbb{CP}^{d-1}$, whose real dimension is $2d - 2 < d^2 - 1$. Therefore,

$$\mathrm{rank}\, F \leq 2d - 2 < d^2 - 1. \tag{S108}$$

For $N = 2$, Secs. S3-S5 prove that weak compatibility excludes full rank. For $N \geq 3$, Eqs. (S101)-(S106) give an explicit compatible, maximizing, fully identifiable probe. Hence, a pure compatible, budget-maximizing, fully identifiable collective probe exists if and only if $N \geq 3$. For mixed probes, covariance domination gives

$$F_Q(\rho) \preccurlyeq 4\,\mathrm{Cov}_\rho(G), \tag{S109}$$

and therefore

$$\mathrm{tr}\, F_Q(\rho) \leq \frac{4N(d-1)(d+N)}{d} \tag{S110}$$

The same scalar cap persists under parameter-independent post-encoding channels by QFIM monotonicity. The pure-state equality classification and the bipartite compatibility-singularity theorem are not asserted for arbitrary mixed probes.

**S13 Fully identifiable large-*d* scaling floor and exchange-symmetric obstruction**

The results in Secs. S13-S15 are supplementary asymptotic observations. They are not used in the exact Fisher-budget identities, the compatibility-identifiability theorem, the maximally entangled boundary classification, the Holevo-divergence result, or the multipartite threshold established in the Letter.

Let $m = d^2 - 1$ be the number of collective $SU(d)$ parameters. For any fully identifiable bipartite probe, $F > 0$. The arithmetic-harmonic mean inequality gives

$$\mathrm{tr}\, F^{-1} \geq \frac{m^2}{\mathrm{tr}\, F}. \tag{S111}$$

Using the Fisher-budget cap Eq. (S16),

$$\mathrm{tr}\, F^{-1} \geq \frac{\left(d^2-1\right)^2}{8(d-1)(d+2)/d}. \tag{S112}$$

Thus,

$$\operatorname{tr} F^{-1} \geq \frac{d(d-1)(d+1)^2}{8(d+2)} = \Omega(d^3). \tag{S113}$$

Define

$$C_*^{\mathrm{SLD}}(d) = \inf_{\substack{\operatorname{Tr}(AA^\dagger)=1 \\ F(A)>0}} \operatorname{tr}\,[F(A)^{-1}]. \tag{S114}$$

Eq. (S113) gives the universal lower bound on $C_*^{\mathrm{SLD}}(d)$. Exchange-symmetric probes cannot reach this cubic floor by approaching the Fisher-budget cap. Indeed, Eq. (S61) shows that the soft antisymmetric-root sectors diverge as $\Delta_*^{-1}$. More globally, exchange-symmetric probes obey a stronger large-$d$ obstruction. For a fully identifiable exchange-symmetric probe, the amplitudes $s_i$ are positive and pairwise distinct. Let the amplitudes $s_i$ be ordered. At least $M_0 = \left\lfloor \frac{d}{2} \right\rfloor$ of them satisfy $s_i \leq \sqrt{\frac{2}{d}}$; otherwise $\sum_i s_i^2 > 1$. These $M_0$ amplitudes lie in an interval of length at most $\sqrt{2/d}$. Let $g_k$ be the consecutive gaps among them. Then $\sum_k g_k \leq \sqrt{\frac{2}{d}}$. By convexity,

$$\sum_k g_k^{-2} \geq \frac{(M_0-1)^3}{(\sum_k g_k)^2} \geq \frac{d}{2}(M_0-1)^3. \tag{S115}$$

Keeping only these consecutive soft-root terms,

$$C_{\mathrm{sym}}^{\mathrm{SLD}} \geq \tfrac{1}{4}\textstyle\sum_k g_k^{-2} \geq \tfrac{d}{8}(M_0-1)^3. \tag{S116}$$

For $d \geq 6$

$$C_{\mathrm{sym}}^{\mathrm{SLD}} \geq \tfrac{d^4}{512}. \tag{S117}$$

Therefore, exchange-symmetric fully identifiable probes have at least quartic SLD cost scaling in this regime. The cubic floor requires exchange breaking.

**S14. Explicit exchange-broken logarithmic upper bound**

For completeness, we record an explicit exchange-broken probe family that realizes the upper side of the supplementary asymptotic bracket. Let $F_d = \frac{1}{\sqrt{d}}\left(\omega^{jk}\right)_{j,k=0}^{d-1}$, where $\omega = e^{\frac{2\pi i}{d}}$, be the normalized discrete Fourier matrix. Define

$$r_j = 1 + \frac{j}{d}, \qquad Z_d = \sum_{j=0}^{d-1} r_j^2, \qquad \Sigma_d = Z_d^{-1/2}\operatorname{diag}(r_0, \dots, r_{d-1}),$$

and take

$$A_d = \Sigma_d \mathcal{F}_d \tag{S118}$$

Since $\mathrm{Tr}\,\Sigma_d^2 = 1$, the probe is normalized. Moreover,

$$d \le Z_d \le 4d. \tag{S119}$$

These inequalities follow from $1 \le r_j < 2$.

**A. Exact hard-soft decomposition**

Define the real-linear map on the real Hilbert space of traceless Hermitian matrices by $\tau(h) = F_d h^\top F_d^\dagger$. Because ${F_d}^\top = F_d$ and $F_d \bar{F}_d = I_d$ the map $\tau$ is an orthogonal involution $\tau^2 = I$. Decompose every traceless Hermitian generator as $h = h_+ + h_-$, $h_\pm = \frac{1}{2}(h \pm \tau h)$, and $\tau h_\pm = \pm h_\pm$. Right-multiplication of the tangent coefficient matrix by $F_d^\dagger$ gives

$$(hA_d + A_d h^\top)F_d^\dagger = h\Sigma_d + \Sigma_d \tau h = \{h_+, \Sigma_d\} + [h_-, \Sigma_d]. \tag{S120}$$

The anticommutator $\{h_+, \Sigma_d\}$ is Hermitian, whereas the commutator $[h_-, \Sigma_d]$ is anti-Hermitian. They are therefore orthogonal in the real Hilbert-Schmidt inner product. The reduced states are $\rho_a = \Sigma_d^2$, $\rho_b = \tau(\Sigma_d^2)$. Therefore, $\rho_a + \rho_b$ belongs to the $\tau = +1$ sector, and consequently $\mathrm{Tr}\,[(\rho_a + \rho_b)h_-] = 0$. For $h_+$, $\mathrm{Tr}\,[(\rho_a + \rho_b)h_+] = 2\,\mathrm{Tr}\,(\Sigma_d^2 h_+) = \langle \Sigma_d, \{h_+, \Sigma_d\}\rangle_{\mathrm{HS}}$. Since $\|\Sigma_d\|_{\mathrm{HS}} = 1$, the complete pure-state QFIM quadratic form is

$$\frac{1}{4} h^\top F(A_d) h = \left\| P_{\Sigma_d^\perp}\{h_+, \Sigma_d\} \right\|_{\mathrm{HS}}^2 + \|[h_-, \Sigma_d]\|_{\mathrm{HS}}^2 \tag{S121}$$

where $P_{\Sigma_d^\perp}$ denotes Hilbert-Schmidt projection orthogonal to $\Sigma_d$. Eq. (S121) includes the complete covariance subtraction. It also proves that the QFIM is exactly block diagonal in the $\tau = \pm 1$ decomposition. We refer to the $\tau = +1$ and $\tau = -1$ subspaces as the hard and soft sectors, respectively.

**B. Uniform lower bound on the hard sector**

Define $A_{\Sigma_d}(x) = \{x, \Sigma_d\}$. Since every eigenvalue of $\Sigma_d$ is at least $Z_d^{-1/2}$, $\left\|A_{\Sigma_d}^{-1}\right\| \le \frac{\sqrt{Z_d}}{2}$. For a traceless $h_+$, write $A_{\Sigma_d}(h_+) = c\Sigma_d + R$, $R \perp \Sigma_d$. The hard-sector contribution in Eq. (S121) is $\|R\|_{\mathrm{HS}}^2$. Since $A_{\Sigma_d}^{-1}(\Sigma_d) = \frac{1}{2} I_d$, we have $h_+ = \frac{c}{2} I_d + A_{\Sigma_d}^{-1}(R)$. Tracelessness gives

$$\frac{cd}{2} = -\mathrm{Tr}\left[A_{\Sigma_d}^{-1}(R)\right]. \tag{S122}$$

Because $A_{\Sigma_d}^{-1}$ is self-adjoint in the Hilbert-Schmidt inner product,

$$\left|\mathrm{Tr}\left[A_{\Sigma_d}^{-1}(R)\right]\right| = \left|\langle A_{\Sigma_d}^{-1}(I_d), R\rangle_{\mathrm{HS}}\right|. \tag{S123}$$

Moreover, $A_{\Sigma_d}^{-1}(I_d) = \frac{1}{2}\Sigma_d^{-1}$ and hence

$$\left\|A_{\Sigma_d}^{-1}(I_d)\right\|_{\mathrm{HS}}^2 = \frac{Z_d}{4}\sum_{j=0}^{d-1}\frac{1}{r_j^2} \leq \frac{dZ_d}{4}. \tag{S124}$$

Therefore, $|c| \leq \sqrt{\frac{Z_d}{d}}\,\|R\|_{\mathrm{HS}}$. Using the inverse-norm bound for $A_{\Sigma_d}$ , $\|h_+\|_{\mathrm{HS}} \leq \frac{|c|\sqrt{d}}{2} + \left\|A_{\Sigma_d}^{-1}(R)\right\|_{\mathrm{HS}} \leq \sqrt{Z_d}\|R\|_{\mathrm{HS}}$. Consequently,

$$4\left\|P_{\Sigma_d^{\perp}}\{h_+,\Sigma_d\}\right\|_{\mathrm{HS}}^2 \geq \frac{4}{Z_d}\|h_+\|_{\mathrm{HS}}^2. \tag{S125}$$

Hence, every hard-sector Fisher eigenvalue is at least $\frac{4}{Z_d} \geq 1/d$. Since the hard-sector dimension is at most $d^2-1$, $\mathrm{tr}\, F_+^{-1} \leq \frac{Z_d}{4}(d^2-1) \leq d^3$

**C. Comparison of the soft sector with the clock-shift Laplacian**

Let $R_d = \mathrm{diag}\,(r_0,\dots,r_{d-1}\,)$, $\Sigma_d = \frac{R_d}{\sqrt{Z_d}}$, and define the clock and shift matrices by $W|j\rangle = \omega^j|j\rangle$, $X|j\rangle = |j+1\rangle$, with indices understood modulo $d$. For any matrix $H$,

$$\|[H,R_d]\|_{\mathrm{HS}}^2 = \sum_{i,j=0}^{d-1}\left|r_i - r_j\right|^2\left|H_{ij}\right|^2. \tag{S126}$$

whereas $\|[H,W]\|_{\mathrm{HS}}^2 = \sum_{i,j=0}^{d-1}\left|\omega^i-\omega^j\right|^2\left|H_{ij}\right|^2$. Since $\left|r_i-r_j\right| = \frac{|i-j|}{d}$ and $\left|\omega^i-\omega^j\right| = 2\left|\sin\frac{\pi(i-j)}{d}\right| \leq \frac{2\pi|i-j|}{d}$, we obtain

$$\|[H,\Sigma_d]\|_{\mathrm{HS}}^2 \geq \frac{1}{4\pi^2 Z_d}\|[H,W]\|_{\mathrm{HS}}^2 \tag{S127}$$

The map $\tau$ reverses products, $\tau(AB) = \tau(B)\tau(A)$ and satisfies $\tau(W) = F_d W F_d^{\dagger} = X^{-1}$.

For $h_-$, orthogonality of $\tau$ and $\tau h_- = -h_-$ imply $\|[h_-,\Sigma_d]\|_{\mathrm{HS}} = \|[h_-,\tau(\Sigma_d)]\|_{\mathrm{HS}}$. Applying Eq. (S127) both before and after transformation by $\tau$, and averaging the resulting lower bounds, gives

$$\|[h_-,\Sigma_d]\|_{\mathrm{HS}}^2 \geq \frac{1}{8\pi^2 Z_d}\left[\|[h_-,W]\|_{\mathrm{HS}}^2 + \|[h_-,X^{-1}]\|_{\mathrm{HS}}^2\right] \tag{S128}$$

Define the positive superoperator $\mathcal{L}_d = \mathrm{ad}_W^{\dagger}\mathrm{ad}_W + \mathrm{ad}_{X^{-1}}^{\dagger}\mathrm{ad}_{X^{-1}}$, where $\mathrm{ad}_Y(H) = [H,Y]$. Then, on the soft sector,

$$F_- \succcurlyeq \frac{1}{2\pi^2 Z_d}\mathcal{L}_d. \tag{S129}$$

This operator inequality controls the generic, line, and self-conjugate modes simultaneously. No separate exceptional-sector estimate is required.

**D. Exact Laplacian spectrum and multiplicity count**

Introduce the normalized displacement matrices $D_{pq} = d^{-1/2} X^p W^q,\ \ p, q \in \mathbb{Z}_d$. They form an orthonormal basis of the complex matrix space. Direct calculation gives $[D_{pq}, W] = (1 - \omega^p) D_{p,q+1}$, and $[D_{pq}, X^{-1}] = (\omega^{-q} - 1) D_{p-1,q}$. Therefore, $\mathcal{L}_d D_{pq} = \mu_{pq} D_{pq}$, where

$$\mu_{pq} = |1 - \omega^p|^2 + |1 - \omega^q|^2 = 4\left[\sin^2\left(\frac{\pi p}{d}\right) + \sin^2\left(\frac{\pi q}{d}\right)\right]. \tag{S130}$$

The only zero eigenvalue is $\mu_{00} = 0$, which corresponds to the identity matrix. The Fourier-transpose involution acts on the displacement basis as $\tau(D_{pq}) = D_{-q,-p}$. Since $\mu_{-q,-p} = \mu_{pq}$, the superoperator $\mathcal{L}_d$ commutes with $\tau$. The identity mode belongs to the $\tau = +1$ sector and is absent from the traceless $\tau = -1$ sector. Hence, $\mathcal{L}_d$, and therefore $F_-$, is strictly positive on the soft sector. Define $p_0 = \min(p, d-p)$, $q_0 = \min(q, d-q)$. For $0 \le p_0, q_0 \le d/2$, $\sin\left(\frac{\pi p_0}{d}\right) \ge \frac{2p_0}{d}$, $\sin\left(\frac{\pi q_0}{d}\right) \ge \frac{2q_0}{d}$. Thus, $\mu_{pq} \ge \frac{16}{d^2}(p_0^2 + q_0^2)$. Because $\mathcal{L}_d$ commutes with $\tau$,

$$\mathrm{tr}\left[\left(\mathcal{L}_{d|\mathcal{H}_-}\right)^{-1}\right] \le \sum_{(p,q)\ne(0,0)} \frac{1}{\mu_{pq}}, \tag{S131}$$

where $\mathcal{H}_-$ denotes the real $\tau = -1$ Hermitian sector. Consequently,

$$\sum_{(p,q)\ne(0,0)} \frac{1}{\mu_{pq}} \le \frac{d^2}{16} \sum_{(p,q)\ne(0,0)} \frac{1}{(p_0^2 + q_0^2)}. \tag{S132}$$

Each pair $(p_0, q_0)$ occurs at most four times. Grouping the first-quadrant lattice points by $r = \max(p_0, q_0)$ gives at most $2r + 1$ points in the $r$-th shell. Hence,

$$\sum_{(p,q)\ne(0,0)} \frac{1}{(p_0^2 + q_0^2)} \le 4 \sum_{r=1}^{\lfloor d/2 \rfloor} \frac{2r+1}{r^2} \le 12(1 + \log d). \tag{S133}$$

It follows that

$$\sum_{(p,q)\ne(0,0)} \frac{1}{\mu_{pq}} \le \frac{3}{4} d^2 (1 + \log d). \tag{S134}$$

Using Eq. (S129) and $Z_d \le 4d$,

$$\mathrm{tr}\, F_-^{-1} \le 2\pi^2 Z_d\, \mathrm{tr}\left[\left(\mathcal{L}_{d|\mathcal{H}_-}\right)^{-1}\right] \le 2\pi^2 Z_d \sum_{(p,q)\ne(0,0)} \frac{1}{\mu_{pq}} \le 6\pi^2 d^3 (1 + \log d). \tag{S135}$$

Combining the hard- and soft-sector estimates gives

$$\operatorname{tr} F(\mathrm{A_d})^{-1} = \operatorname{tr} F_+^{-1} + \operatorname{tr} F_-^{-1} \leq d^3 + 6\pi^2 d^3(1+\log d) \\ \leq (1+6\pi^2)d^3(1+\log d). \tag{S136}$$

Specifically,

$$F(A_d) > 0. \tag{S137}$$

for every $d \geq 2$, and

$$\operatorname{tr} F(A_d)^{-1} = O(d^3 \log d) \tag{S138}$$

Combining Eq. (S136) with the universal lower bound in Sec. S13 gives the unconditional bracket

$$\frac{d(d-1)(d+1)^2}{8(d+2)} \leq C_*^{\mathrm{SLD}}(d) \leq (1+6\pi^2)d^3(1+\log d). \tag{S139}$$

Hence,

$$C_*^{\mathrm{SLD}}(d) = \Omega(d^3),\ C_*^{\mathrm{SLD}}(d) = O(d^3 \log d). \tag{S140}$$

The lower bound is universal, whereas the upper bound is realized by the explicit affine-profile Fourier-frame family in Eq. (S118).

**S15. Open problem: closing the logarithmic gap**

The bracket in Eq. (S139) is unconditional. Whether the logarithmic factor can be removed remains open. It would be sufficient, though not necessary, to construct a fully identifiable family satisfying $F(A_d) \succcurlyeq \frac{c}{d} I_{\mathfrak{su}(d)}$ with a constant $c > 0$ independent of $d$. Such a family would obey

$$\operatorname{tr} F(A_d)^{-1} = O(d^3) \tag{S141}$$

No such construction is proved here, and no result from this section is used in the exact theorems of the Letter.

**S16. Scope of claims**

The exact bipartite Fisher-budget identities are Eqs. (S11), (S15), and (S17). The kernel characterization is Eq. (S19). The universal compatibility-induced blindness theorem is Eq. (S49), which proves $\Omega = 0 \Longrightarrow \operatorname{corank} F \geq \left\lfloor \frac{d}{2} \right\rfloor$ for every pure bipartite probe. The preliminary binary singularity statement is Eq. (S44). The complete boundary-spectrum theorem, Eqs. (S31) and (S32), applies to maximally entangled bipartite probes. The full spectrum of $V = U\overline{U}$, including its eigen phases and multiplicities, fixes the QFIM eigenvalues and their multiplicities. It does not by itself fix the laboratory orientation of the Fisher eigenspaces; that orientation depends on the antiunitary frame $J = UC$.

The projector identity Eq. (S35) applies to the extreme strata $V = \pm I_d$. The $V = -I_d$ stratum exists only for even $d$. The Fisher-budget maximum occurs at $V = +I_d$. The maximizing symmetric-space projector is Eq. (S22), corresponding to Eq. (6) of the main text, while the maximizing spectrum is Eq. (S36), corresponding to Eq. (10).

The exchange-symmetric Holevo formula Eq. (S58) is exact for regular exchange-symmetric probes with $\lambda_i > 0, \lambda_i \neq \lambda_j$. Degenerate spectra are interpreted through a limiting procedure and produce divergent costs. The formula is not claimed to determine the globally optimal probe over all bipartite probe families.

The deficit-controlled bounds in Eqs. (S63) and (S69) together give Eq. (13) of the main text. The qutrit comparison in Sec. S11 concerns two paths with the same Schmidt spectrum but different relative Schmidt frames. The antiunitary class fixes the number of soft channels, while the divergence amplitude remains path dependent.

The multipartite scalar budget law extends to arbitrary $N$, but the compatibility-identifiability exclusion is specific to $N = 2$. Eqs. (S101) and (S104) together give Eq. (14) of the main text. The generalized GHZ family is an explicit construction proving that the obstruction can be evaded for $N \geq 3$; it is not claimed to be the unique multipartite maximizing family.

For the identity-weight SLD cost optimized over fully identifiable pure bipartite probes, Secs. S13 and S14 establish the unconditional bracket in Eq. (S139). The lower bound is universal, while the upper bound is realized by the explicit affine-profile Fourier-frame family $A_d = \Sigma_d \mathcal{F}_d$ defined in Eq. (S118). Fully identifiable exchange-symmetric probes obey an order-$d^4$ lower bound. Removal of the logarithmic factor remains open; no unconditional $O(d^3)$ upper bound is claimed.

**S17. Supplemental References**

[S1] R. A. Horn and C. R. Johnson, *Matrix Analysis*, 2nd ed. (Cambridge University Press, Cambridge, 2013).

[S2] R. A. Horn and V. V. Sergeichuk, “Canonical forms for unitary congruence and *congruence,” *Linear Multilinear Algebra* **57**, 777-815 (2009).

[S3] F. J. Dyson, “The threefold way. Algebraic structure of symmetry groups and ensembles in quantum mechanics,” *J. Math. Phys.* **3**, 1199-1215 (1962).

[S4] A. S. Holevo, *Probabilistic and Statistical Aspects of Quantum Theory*, 2nd ed. (Edizioni della Normale, Pisa, 2011).

[S5] K. Matsumoto, “A new approach to the Cramér-Rao-type bound of the pure-state model,” *J. Phys. A* **35**, 3111-3123 (2002).
[S6] J. Friel, P. Palittapongarnpim, F. Albarelli, and A. Datta, “Attainability of the Holevo–Cramér-Rao bound for two-qubit 3D magnetometry,” arXiv:2008.01502 [quant-ph] (2020).
[S7] L. Pezzè, A. Smerzi, M. K. Oberthaler, R. Schmied, and P. Treutlein, “Quantum metrology with nonclassical states of atomic ensembles,” *Rev. Mod. Phys.* **90**, 035005 (2018).